\documentclass[a4paper,11pt]{article}
\usepackage{jinstpub} 

\usepackage{url}
\usepackage{color}

\arxivnumber{2209.08002}

\begin{document}

%\begin{frontmatter}



\title{Characterization of XIA UltraLo-1800 Response to Measuring Charged Samples}

\author{Joshua Ange}%\corref{cor1}}
%\ead{joshange@gmail.com }
\author{Robert Calkins}
\author{Andrew Posada}
\emailAdd{joshange@gmail.com}
%\address{ Department of Physics, Southern Methodist University, Dallas, TX 75275, USA} 

%Why is this broken? 
\affiliation{
organization={Department of Physics, Southern Methodist University},
%%%            %addressline={}, 
            city={Dallas},
            state={TX},
            postcode={75025}, 
country={USA}}

%\begin{abstract}
\abstract{
Commercial alpha counters are used in science and industry applications to screen materials for surface radon progeny contamination. In this paper, we characterize an XIA UltraLo-1800, an ionization drift alpha counter, and study the response to embedded charge in polyethylene sample measurements. We show that modeling such effects is possible in a Geant4-based simulation framework and attempt to derive corrections. This paper also demonstrates the effectiveness of the use of an anti-static fan to eliminate the embedded charge and recover a 97.73\% alpha detection efficiency in the alpha counter.
}
%\end{abstract}

%\includegraphics[width=0.75\columnwidth]{grabs}


%\begin{keyword}
%alpha assay \sep simulation \sep electrostatic potential \sep XIA UltraLo-1800 \sep polyethylene \sep backgrounds

%\end{keyword}

%\end{frontmatter}
\maketitle
\flushbottom

\section{Introduction}
\begin{figure}
    \centering
    \includegraphics[width=0.75\columnwidth,page=1]{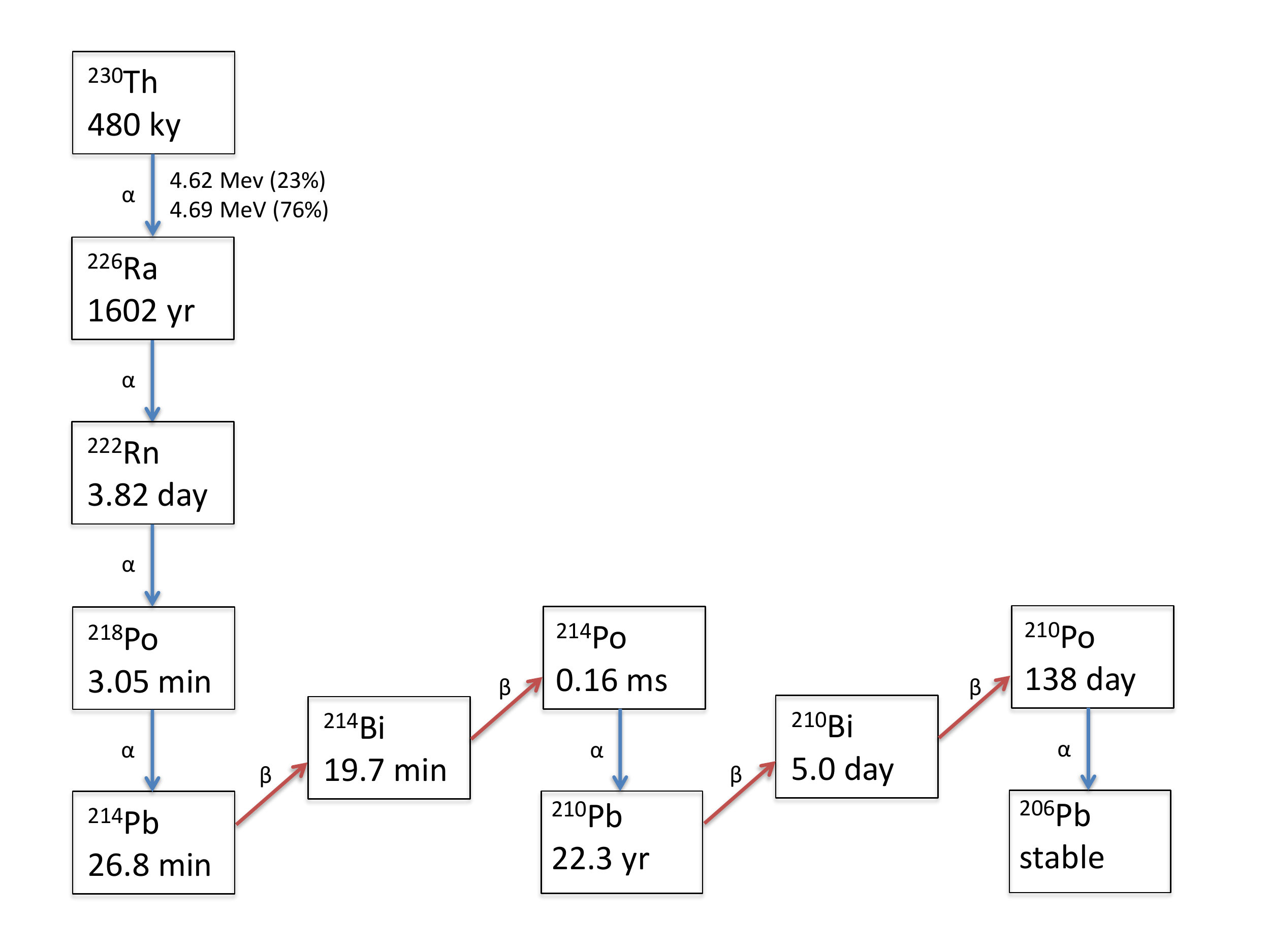}
    \caption{Diagram of $^{230}$Th decay chain.}
    \label{Th230_chain}
\end{figure}

Dark matter is a hidden, unseen form of matter believed to comprise the majority of matter in the universe and influence galaxy rotation, cluster motion, and the large-scale structure of the universe \cite{Planck:2018vyg}. The nature of dark matter is largely unknown and has become a subject of increased study regarding its precise composition and characteristics. Low background experiments, such as SuperCDMS SNOLAB \cite{SuperCDMS:2016wui}, utilize sensitive particle detectors to observe dark matter interactions with silicon and germanium crystals. To directly detect dark matter, these sensitive detectors are placed in low-background, underground environments shielded from cosmic rays and other noise. Due to their high sensitivity and the low number of expected signal events, it becomes vital that the material surfaces involved in the detector are radiopure, which becomes difficult due to the high radon levels often present in underground laboratories. Radon daughters in the air can plate-out, causing $^{210}$Pb atoms to accumulate on surfaces. Such radon progeny will further decay and produce alpha particles (see Fig. \ref{Th230_chain}) that imitate the expected signals of dark matter interactions \cite{Bunker:2020sxw}.

It was observed in Ref. \citenum{STEIN201892} that there was a notable variation in the observed accumulation rate of $^{210}$Pb on polyethylene when compared to copper samples. It was suggested that the discrepancy in plate-out rates may be due to static charge that develops on non-conducting surfaces \cite{STEIN201892}. Furthermore, it has been observed in Ref. \citenum{gordon_rodbell_murray_sri-jayantha_mcnally_2015} that measured alpha rates and energy are significantly distorted by the polarity and magnitude of such charge.

In this article, we describe our method of recommissioning an XIA UltraLo-1800 Alpha Particle Counter \cite{XIA} (XIA) and our approach to examining and simulating the response to alpha particles produced by a $^{230}$Th radioactive sample placed on polyethylene panels with embedded charge.
In Section \ref{Recommissioning}, we describe the recommissioning and re-calibration of the XIA. Section \ref{ChargedPoly} details the methodology and findings of our experiments with a series of charged polyethylene panels. Our simulation-based studies of the impact of charged polyethylene are presented in Section \ref{Simulation}. Section \ref{Antistatic} describes the use of an anti-static fan to mitigate charge effects. The paper concludes with a discussion of results and outlook for future research.

\begin{figure}[ht]
    \centering
    \includegraphics[width=0.75\columnwidth]{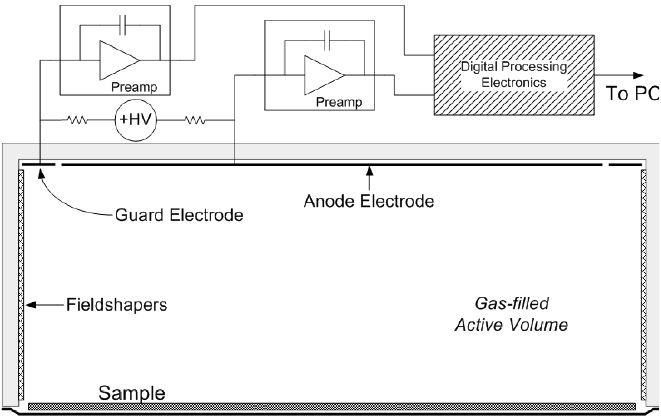}
    \caption{Schematic overview of the UlraLo-1800 chamber and electronics from Ref. \citenum{XIA}.}
    \label{XIA}
\end{figure}

\subsection{The XIA UltraLo-1800}
The XIA UltraLo-1800 Alpha Particle Counter, present in the clean room in the LUMINA laboratory at Southern Methodist University, measures alpha particles originating from radioactive sources. As illustrated in Fig. \ref{XIA}, the XIA’s inner chamber is primarily comprised of a grounded electrode on which a sample can be placed and two positive electrodes at 1100 V (an anode electrode and veto/guard electrode). The anode can be configured to a square ‘full’ anode of 1800 cm$^2$ or circular ‘wafer’ anode of 707 cm$^2$ counting area. For all measurements described in this paper, the ‘wafer’ anode configuration was used. Field shapers along the sides of the chamber are used to maintain parallel electric field lines throughout the argon drift volume. When a radioactive sample is placed in the chamber, emitted alpha particles ionize the argon gas, resulting in an ionization electron track. The freed electrons drift along the field lines and induce a signal in the anode \cite{Shockley,Ramo}. This time-induced charge is filtered, processed and integrated into a digitized pulse. Pulse shape discrimination (PSD) is used to discriminate between alpha particles that originate close to the sample tray versus backgrounds signals, which can originate from in-air radon decay, cosmic rays, or chamber contamination \cite{XIA}

When the sample door is opened, the measurement chamber of the XIA is exposed to moisture and lab air that can affect the drift velocity of electrons. As such, each measurement period following a tray opening was accompanied by a 1-hour purge period with an increased argon flow rate. Additionally, the first six hours of data following a purge were removed from the analysis since it took additional time for the counting gas to stabilize \cite{Bunker:2020sxw}.

\section{Recommissioning the Alpha Particle Counter}
\label{Recommissioning}
The XIA present in the LUMINA laboratory was temporarily decommissioned toward the end of 2019 for roughly 27 months. At the start of this period, no gas flowed into the machine (and it therefore picked up backgrounds from radon exposure from the lab air), but it was purged with N$_2$ boil-off gas for the majority of the decommissioned period in order to prevent further contamination. Upon recommissioning the XIA, we performed calibration tests with the $^{230}$Th sample used in this study and sought additional confirmation for calibration with incandescent gas lantern mantles containing thorium nitrate.

\begin{figure}[ht]
    \centering
    \includegraphics[width=0.75\columnwidth]{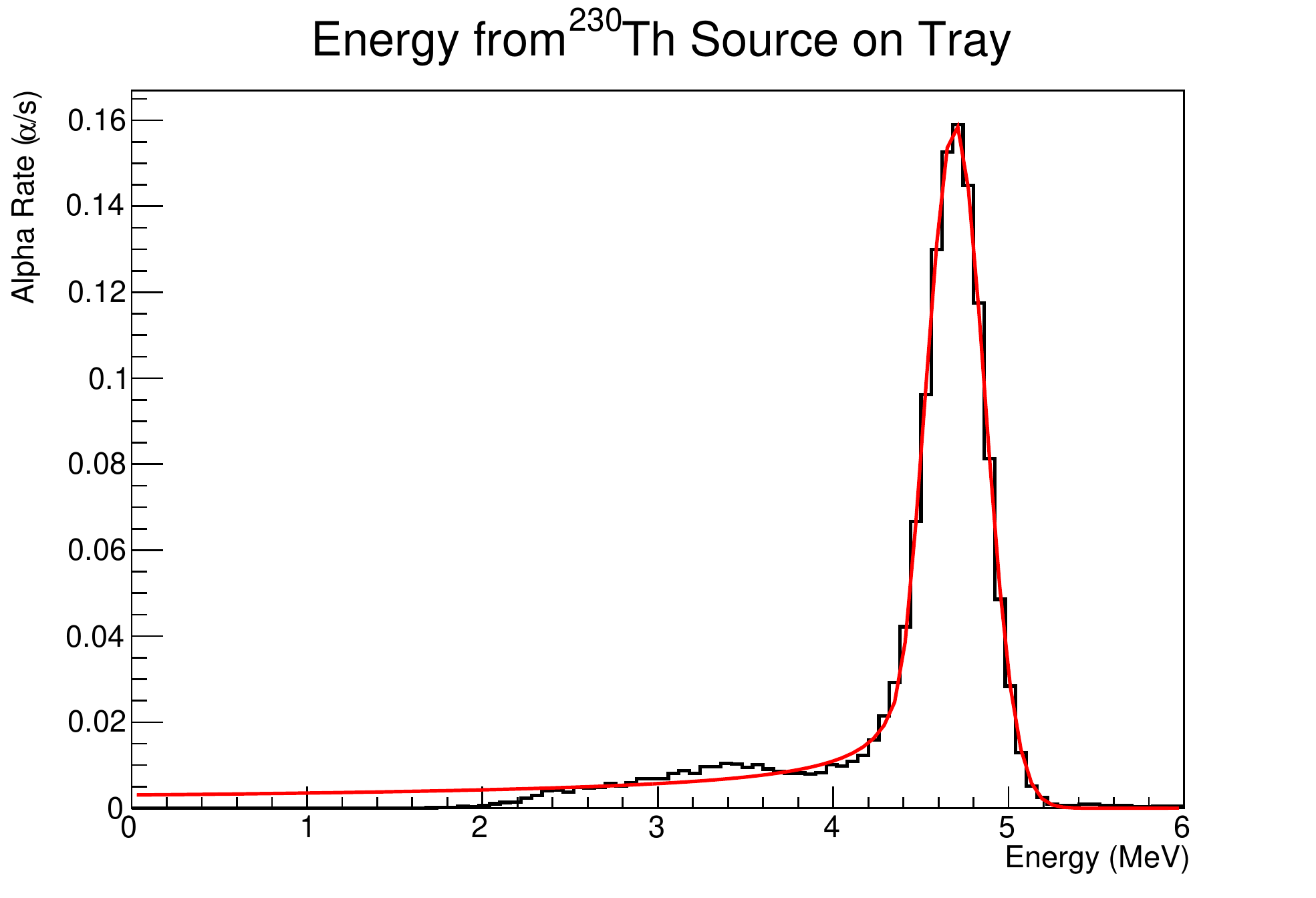}
    \caption{Energy distribution of alpha particles exiting $^{230}$Th sample placed directly on tray (black) and its associated Crystal Ball fit (red). }
    \label{trayEnergy}
\end{figure}

\subsection{Thorium Source Calibration}
The primary radioactive sample involved in these studies was a cylindrical $^{230}$Th disc with a diameter of 1 in and thickness of 2 mm. %Its prevalence in the involved research makes it an appropriate standard to ensure the accuracy of the calibration of the XIA. 
The $^{230}$Th decay chain, as seen in Fig. \ref{Th230_chain}, produces alphas of known energy quantities from 4.6 to 4.7 MeV (23\% 4.62 MeV and 76\% 4.69 MeV) \cite{AKOVALI1996433}.

The $^{230}$Th source disc was placed directly on the conductive Teflon-covered stainless-steel tray of the XIA. The energy of the alpha particle spectrum could be fit by a Gaussian distribution with an additional low energy tail due to $^{230}$Th implantation. A Crystal Ball function was used to fit the distribution \cite{Gaiser:1982yw}. Since the shape of the implantation tail is unknown and not properly modeled by a simple exponential, the uncertainty was artificially inflated to diminish the impact of the tail on the overall fit to the peak, as shown in Fig. \ref{trayEnergy}. The peak value of the exiting energy of alpha particles was measured at 4.69 $\pm$ 0.17 MeV. Additionally, the observed alpha production rate of 1.422 $\pm$ 0.006 $\alpha$/s was consistent with the disc's calibrated alpha emission rate of 1.44 $\pm$ 0.06 $\alpha$/s.

\begin{figure}[ht]
    \centering
    \includegraphics[width=0.75\columnwidth,page=2]{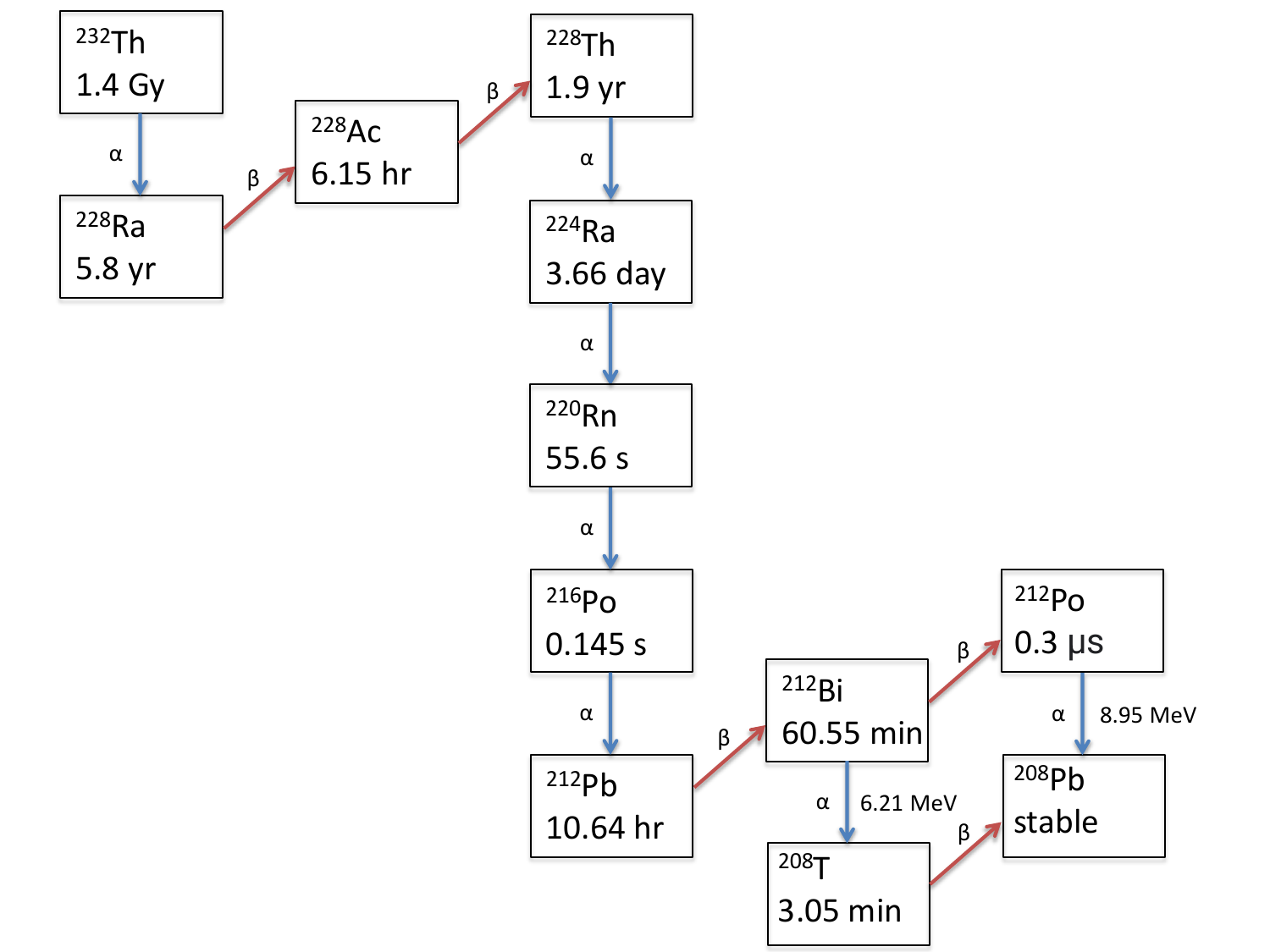}
    \caption{Diagram of $^{232}$Th decay chain.}
    \label{Th232_chain}
\end{figure}

\begin{figure}[ht]
    \centering
    \includegraphics[width=0.75\columnwidth]{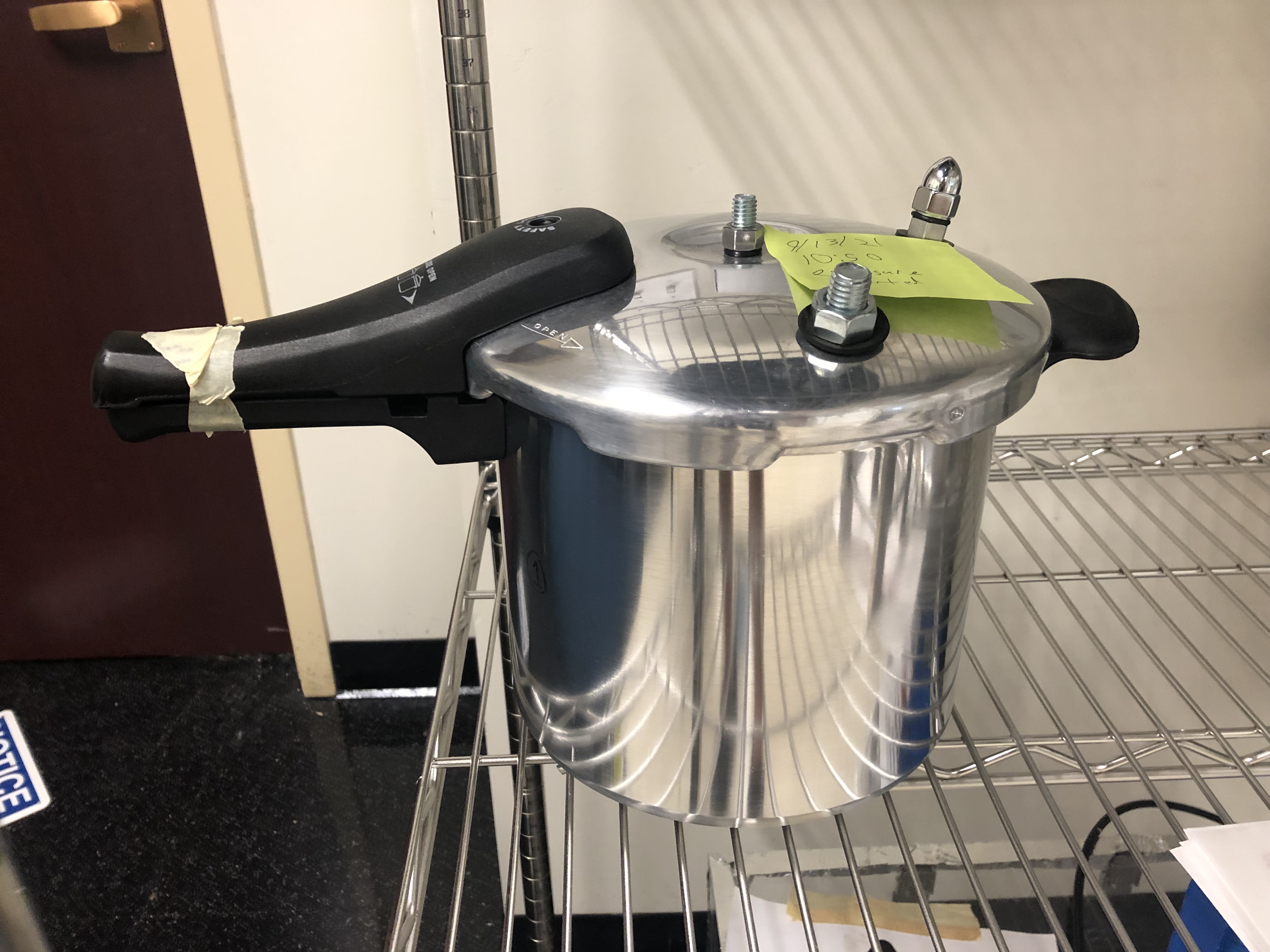}
    \includegraphics[width=0.75\columnwidth]{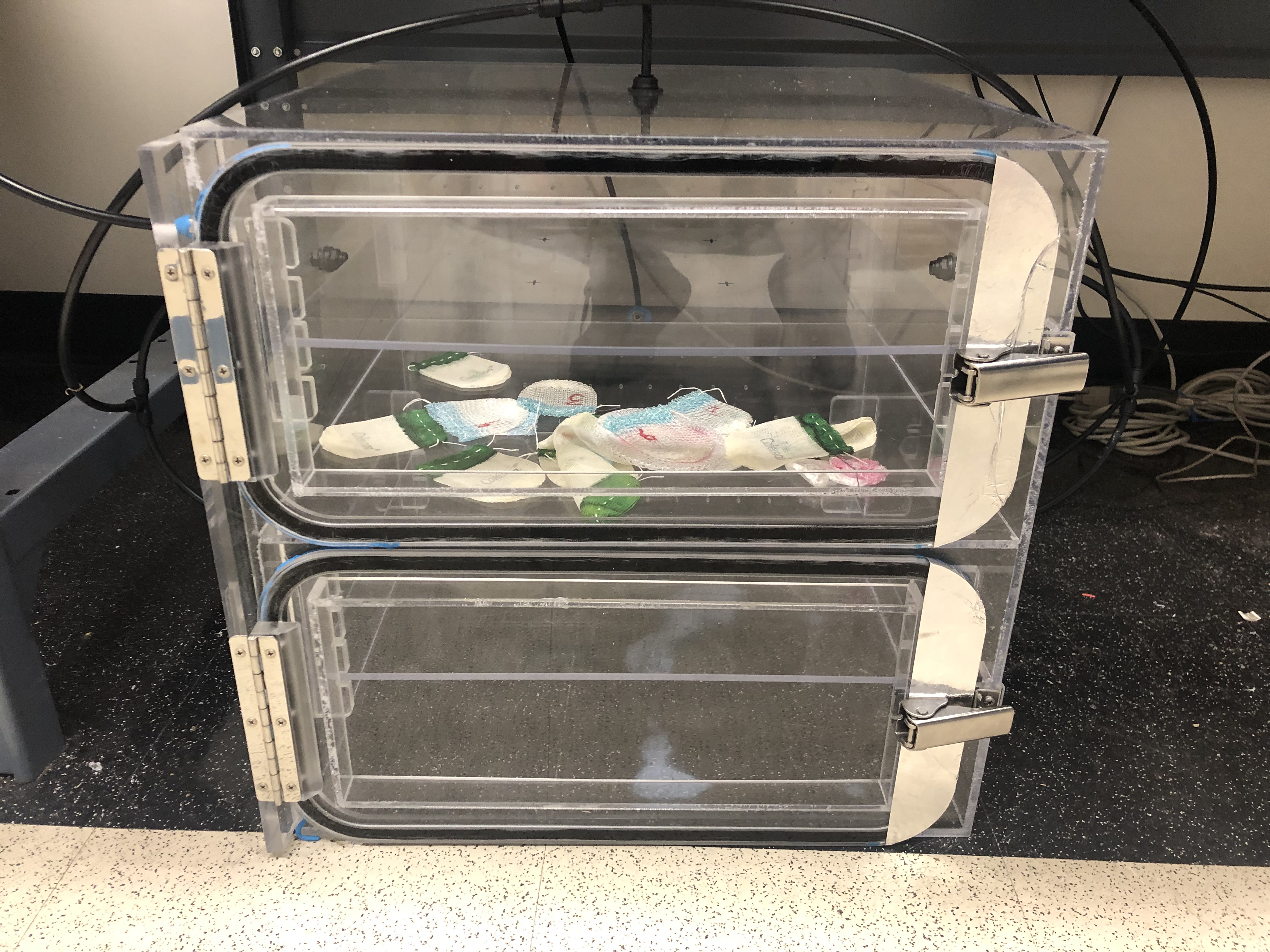}
    \caption{(Top) Photo of pressure cooker modified to serve as exposure chamber for copper plate. (Bottom) Photo of purge box with thorium-containing lantern mantles inside.}
    \label{exposure}
\end{figure}

\subsection{Thorium Lantern Mantle Calibration}
"Welsbach" gas lantern mantles often used for camping contain high levels of thorium, which emit their characteristic glow upon the burning of the mantles' fuel. As a result of this usage, significant levels of thorium daughters become released by the mantle and accumulate on nearby surfaces. A 10 cm×10 cm×3 mm copper square was placed in a sealed pressure cooker with the camping mantles. 
From the accumulation of thorium daughters, the alpha emissivity of this copper square was expected to align with the data from the $^{232}$Th chain in Fig. \ref{Th232_chain}, serving as an additional energy calibration point for the XIA. A similar process was performed with a polyethylene panel of dimensions 24 in×24 in×3/16 in placed within a purge box alongside the thorium-containing lantern mantles (see Fig. \ref{exposure}) and was expected to produce similar results.

\begin{figure}[ht]
    \centering
    \includegraphics[width=0.495\columnwidth]{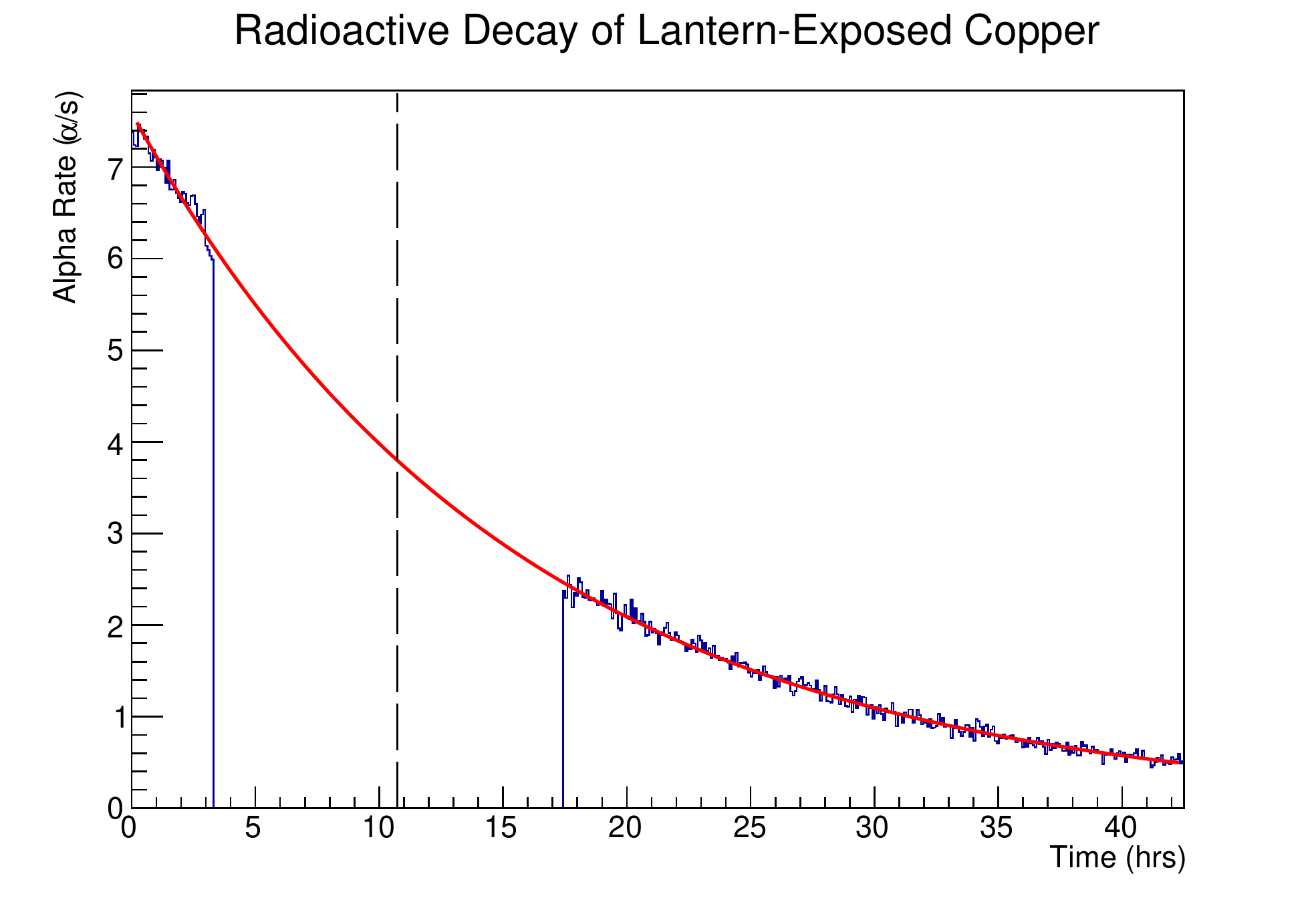}
    \includegraphics[width=0.495\columnwidth]{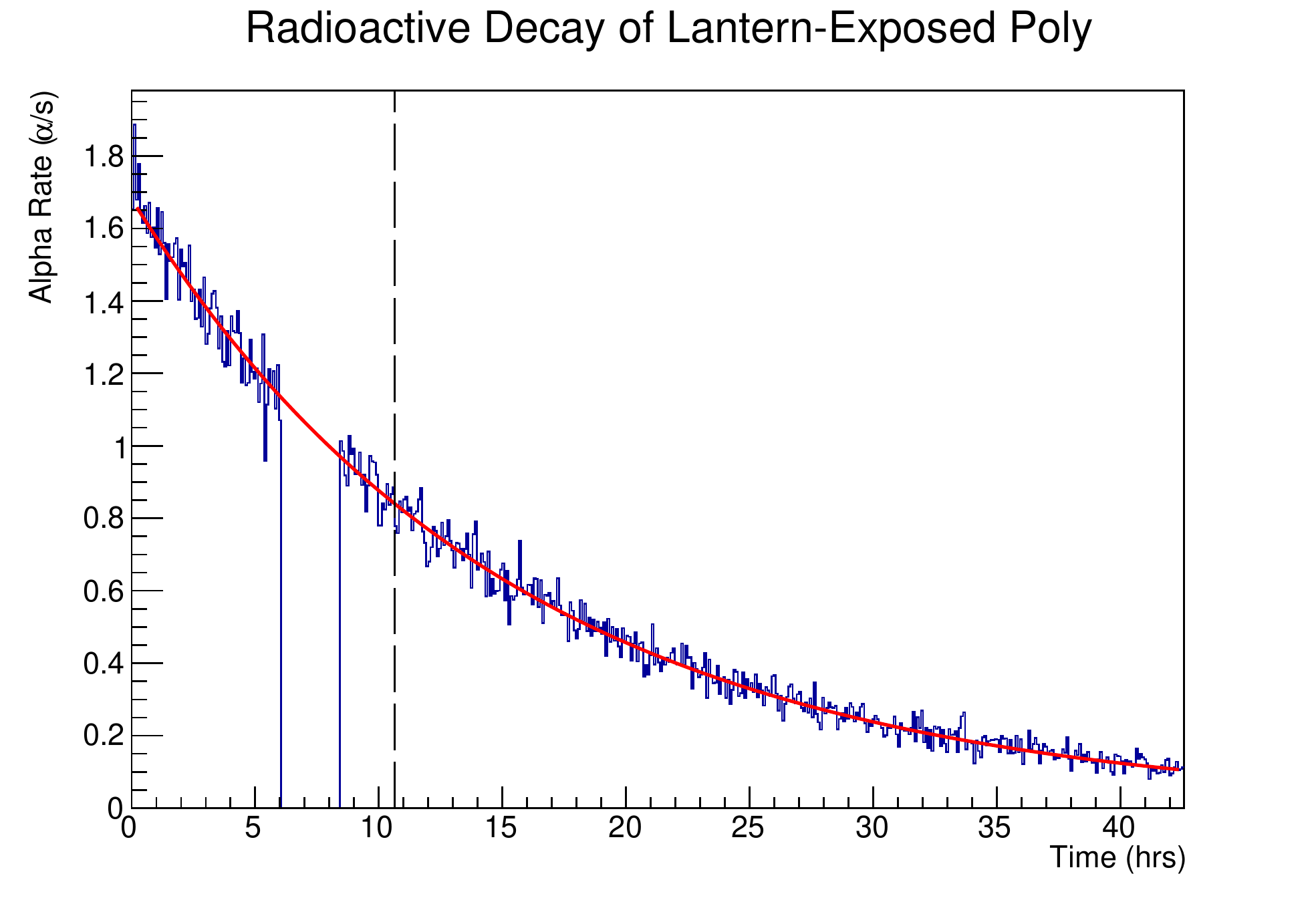}
    \caption{Graphs of radioactive decay with markings at half-life for lantern-exposed copper and polyethylene samples.}
    \label{decay}
\end{figure}

\begin{figure}[ht]
    \centering
    \includegraphics[width=0.495\columnwidth]{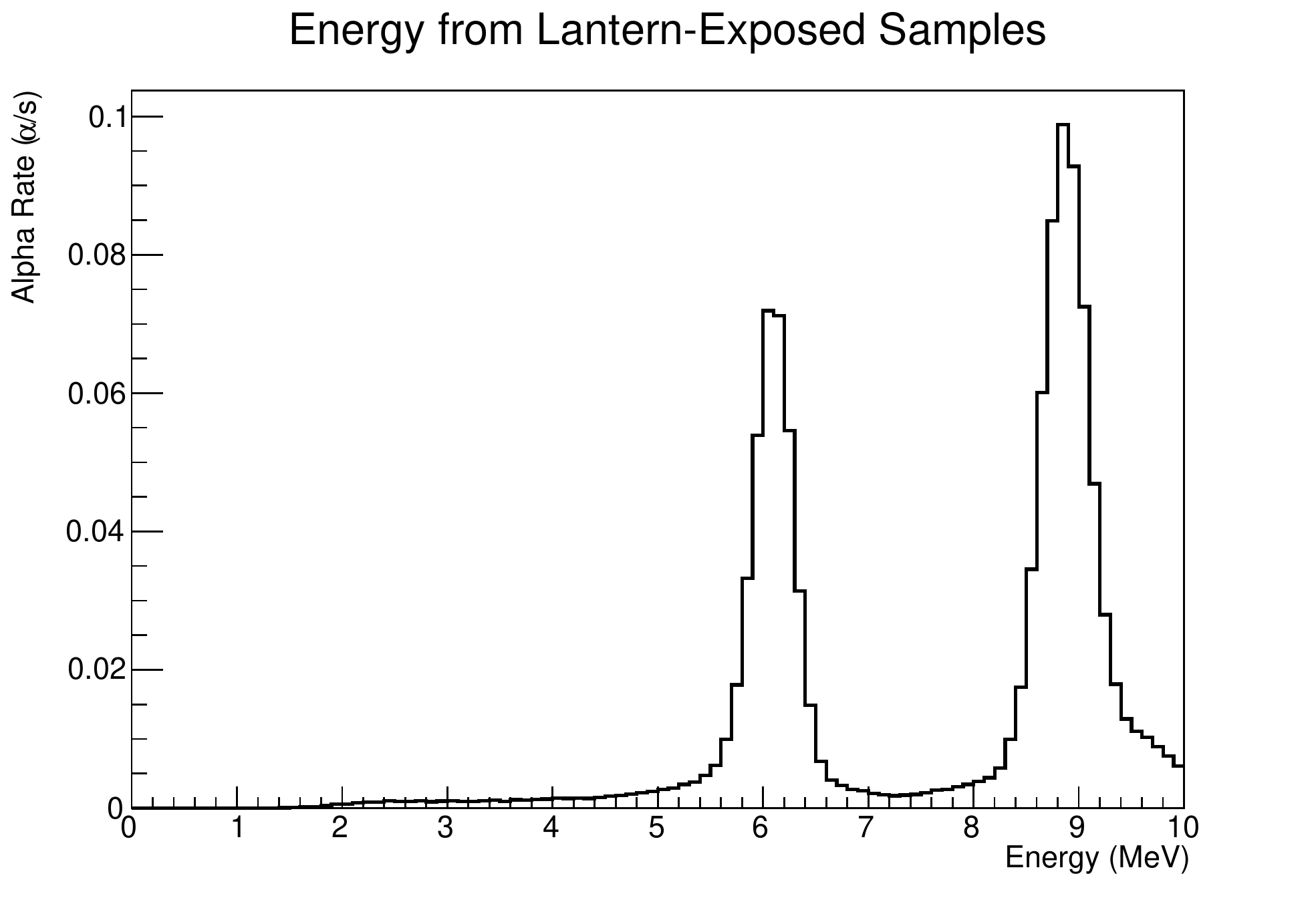}
    \caption{Energy distribution of alpha particles exiting thorium lantern-exposed samples.}
    \label{dualpeaks}
\end{figure}

By fitting the half-life for both the exposed copper and exposed polyethylene (see Fig. \ref{decay}), we obtained values consistent with $^{212}$Pb literature \cite{NNDC}. Additionally, the energy of exiting alphas for both samples (see Fig. \ref{dualpeaks}) aligned with the expected ‘dual peaks’ of alpha energies produced by the decay of $^{212}$Pb daughters (6.21 MeV and 8.95 MeV from $^{212}$Bi and $^{212}$Po, respectively). 

The consistency between the experimental measurements from the newly recommissioned XIA and the calibrated data suggests that the XIA can be stored for extended periods under N$_2$ purge and can be expected to behave normally afterward. No adjustment was needed to re-calibrate the instrument from its calibration prior to storage. 

\section{Series of Charged Polyethylene Samples}
\label{ChargedPoly}

\subsection{Methodology}
The experiment was located in the clean room in the LUMINA laboratory in the Southern Methodist University Physics Department. Three polyethylene panels were used, each of dimensions 24 in×24 in×3/16 in. To avoid contamination, samples placed in the XIA were handled as little as possible and were kept in the clean room (and purge box, when possible). Panels were selected based on a cursory measurement of their electrostatic potentials to experiment with a wide range of charge values.

\begin{figure}[ht]
    \centering
    \includegraphics[width=0.75\columnwidth]{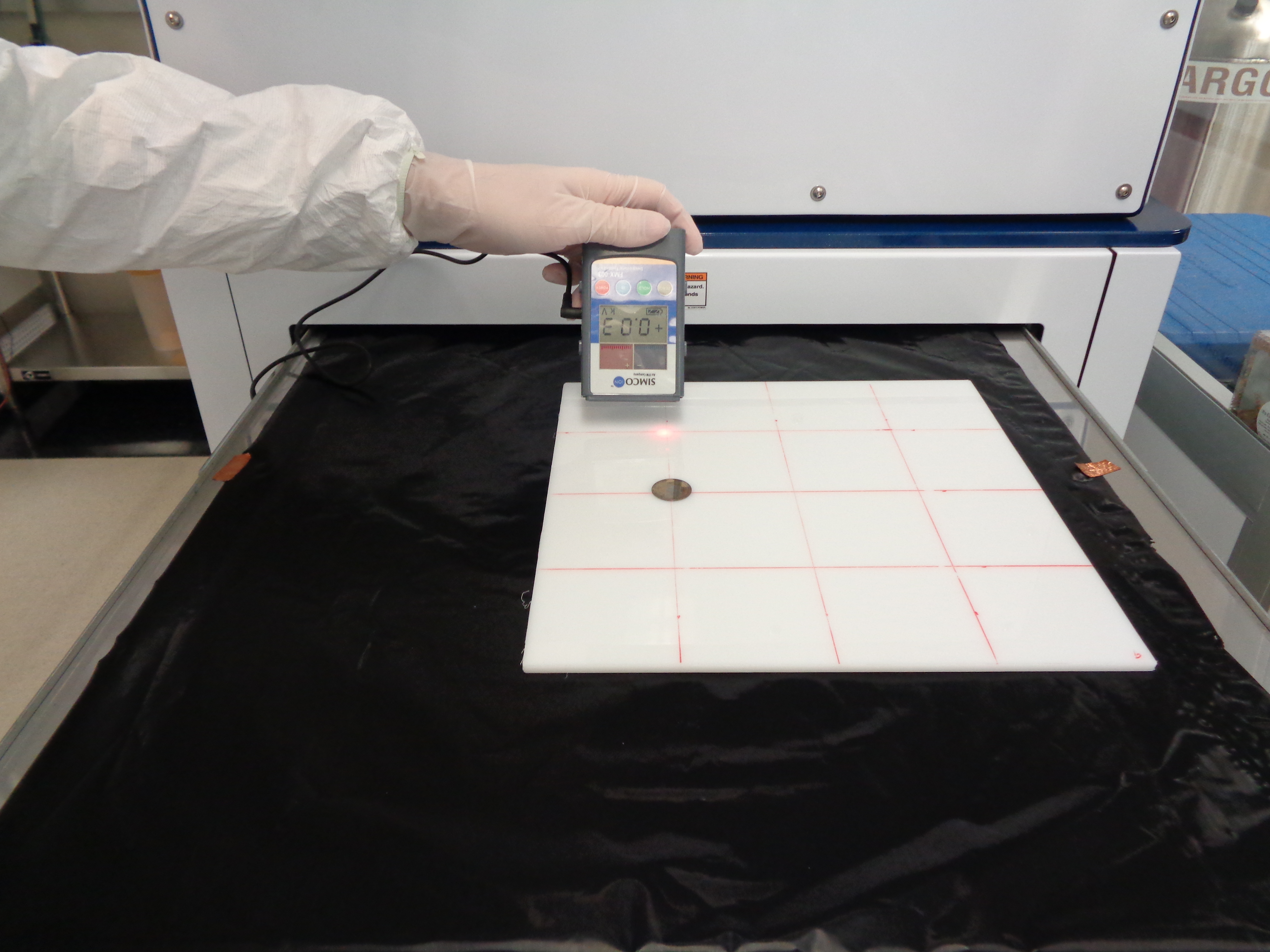}
    \caption{Photo of experimental setup of charged polyethylene on open, Teflon-covered tray of XIA UltraLo-1800 with $^{230}$Th sample on ‘Point 4’ and hand-held voltmeter being used to measure ‘Point 1.’ 4×4 grid can be seen marked in red.}
    \label{experimentalSetup}
\end{figure}

Each polyethylene panel was marked with a 4×4 grid, forming 3 in×3 in squares on its surface. The intersection points of the lines forming this grid served as the primary measurement locations in this study. For each measurement period  (or ‘run’) in the XIA, a point on the polyethylene was chosen for the radioactive $^{230}$Th sample to be placed upon. The radioactive sample and polyethylene panel were moved along the XIA tray such that the $^{230}$Th source was always centered below the anode for highest detection efficiency and to remove any position dependence from the XIA response. 
The electrostatic potential above the polyethylene was measured at the predefined points and scanned at the intervening locations for values of significant deviations. 
We used a Trek 520 Series Hand-held Electrostatic Voltmeter and an FMX 003 Hand-held Electrostatic Voltmeter, which have accuracies of $\pm$ 5\% within a range of 0 to $\pm$ 2 kV and $\pm$ 10\% within a range of 0 to $\pm$ 20 kV, respectively \cite{trek520,fmx003}.

\begin{figure}[ht]
    \centering
    \includegraphics[width=0.75\columnwidth]{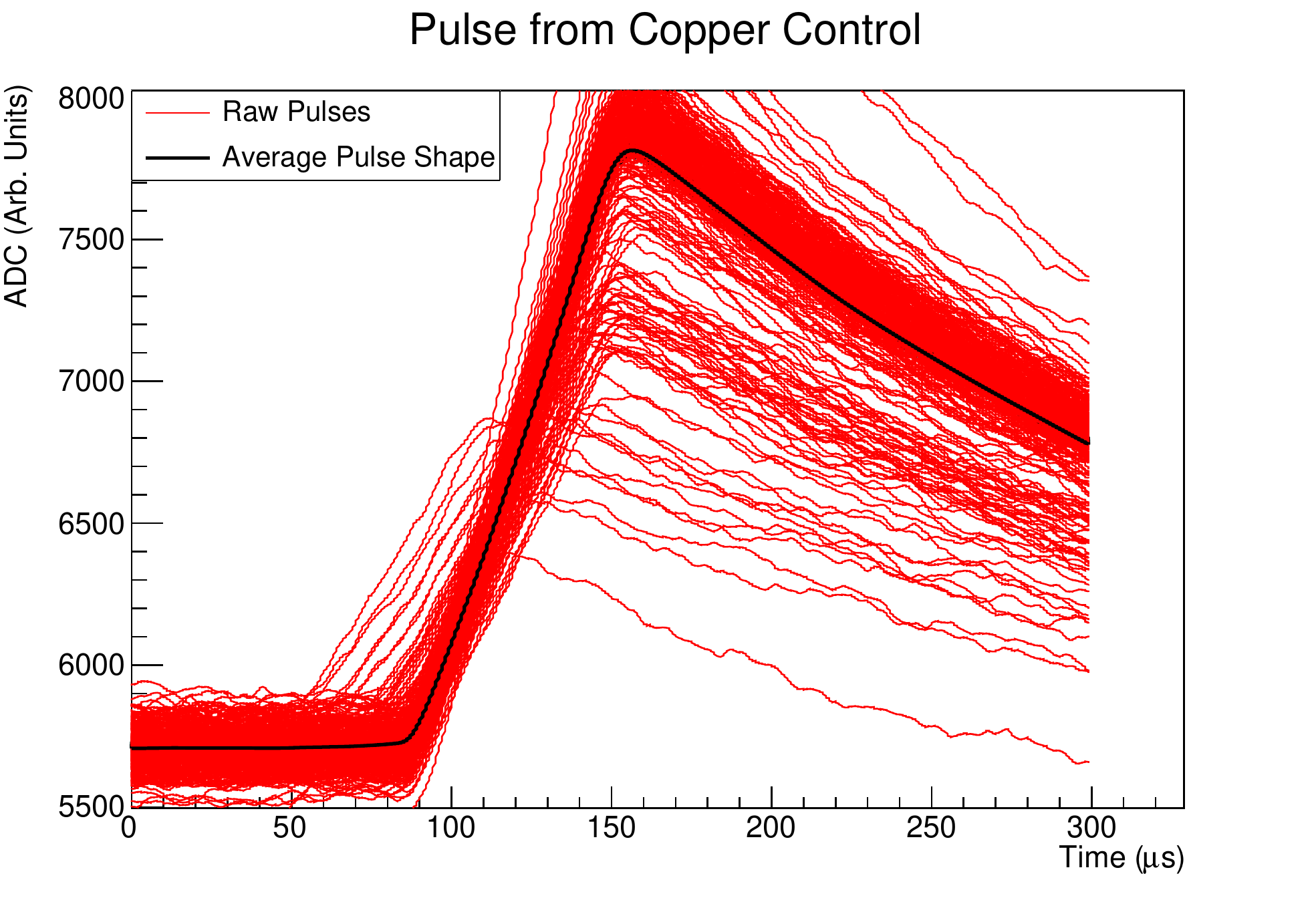}
    \caption{A sampling of 300 pulses of alpha particles from the copper control and the average pulse shape for all alpha particles.}
    \label{averagePulses}
\end{figure}

‘Rough’ and ‘fine’ grid runs were performed. Rough runs involved measurement at each of the initial nine points. Furthermore, when the sample was not located at the center point, such as in Fig. \ref{experimentalSetup}, we measured points along the edge of the polyethylene at 3 in intervals. 
These measurements along the edge and corner positions were an addition to the methodology during data collection so is not present in the earlier runs.
Omission of these values did not make a significant impact on results, likely due to these points’ close proximity to the grounded anode. Additionally, there was no significant difference in electrostatic potential measurements at the location of the $^{230}$Th source when it was removed from the polyethylene. As the charges on the polyethylene surface could vary due to the radiation exposure (and exposure to air during electrostatic potential measurement), measurements were taken immediately before and after each run to limit the impact of drifting charge over the course of measurement and analysis. We estimate a drift of charge in the polyethylene of $\sim$0.44 V/m outside the electric field and therefore $\sim$9 V during the duration of potential measurement, not significant enough to make a significant impact on findings.

\subsection{‘Rough’ Polyethylene Measurements}
\begin{figure}[ht]
    \centering
    \includegraphics[width=0.75\columnwidth]{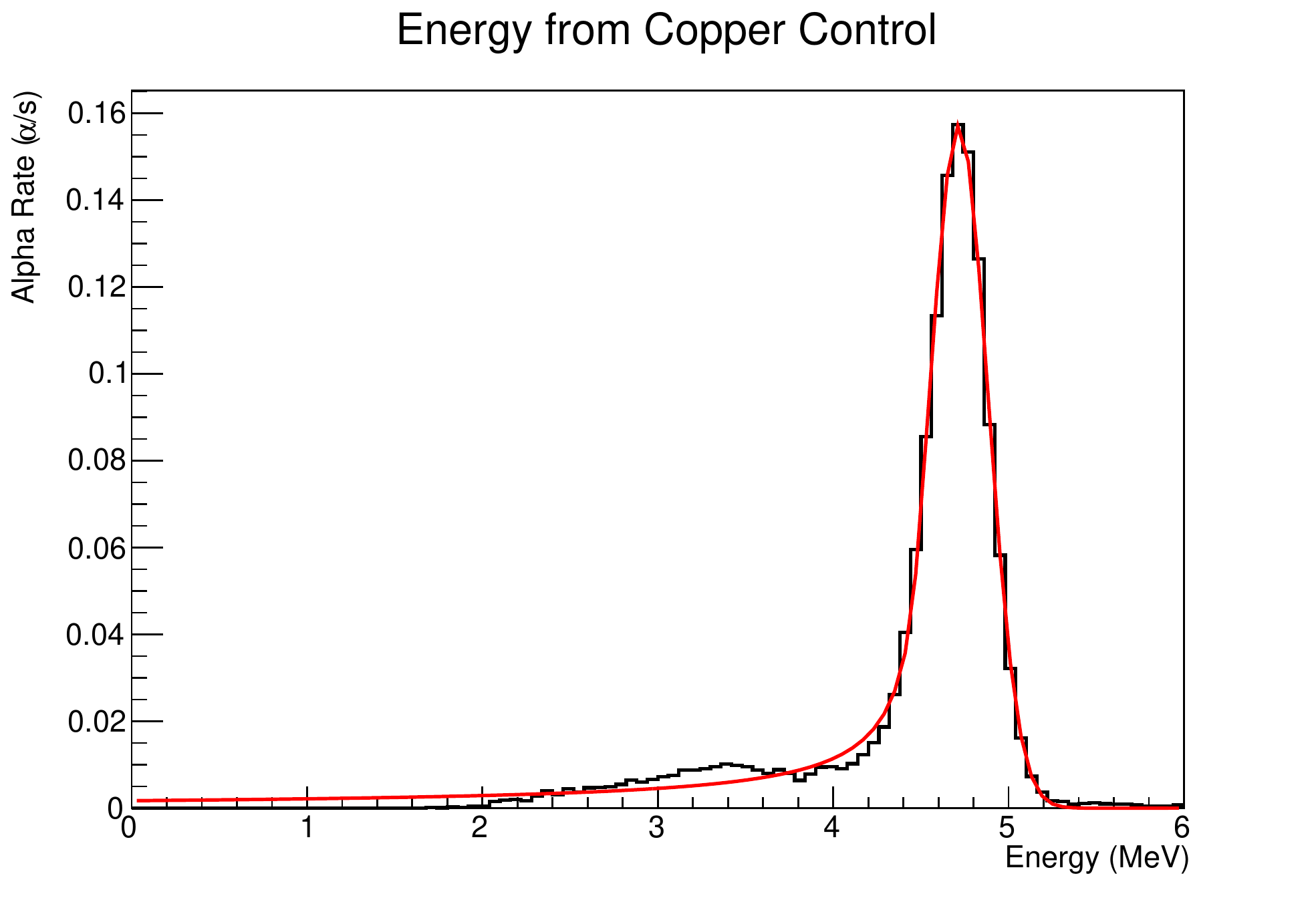}
    \caption{Energy distribution of alpha particles exiting $^{230}$Th sample placed on copper plates.}
    \label{copperEnergy}
\end{figure}

Two copper plates of dimensions 24 in×12 in×2/16 in were placed side-by-side to be used as a control for alpha measurements, as there would be no static charge to influence the electric field. The thickness of the plates is 1/16 in less than that of the polyethylene, so a slight discrepancy could be expected (largely a slightly longer risetime and therefore the slightly higher pulse amplitude of runs with copper), but it was not a large enough effect to invalidate them as controls. 
A sample of pulses identified as alpha particles produced over the course of a sixteen-hour run with the copper control (bearing in mind a six hour cut of the initial data following the purge) can seen in Fig. \ref{averagePulses}. The average pulse shape was primarily used in analysis, as it smoothed baseline noise and made us less sensitive to statistical variations.
For the control run, we measured an alpha emission rate of 1.412 $\pm$ 0.007 $\alpha$/s. Additionally, the energy scale of alpha particles peaked at 4.71 $\pm$ 0.18 MeV (see Fig. \ref{copperEnergy}).

\begin{figure}[ht]
    \centering
    \includegraphics[width=0.9\columnwidth]{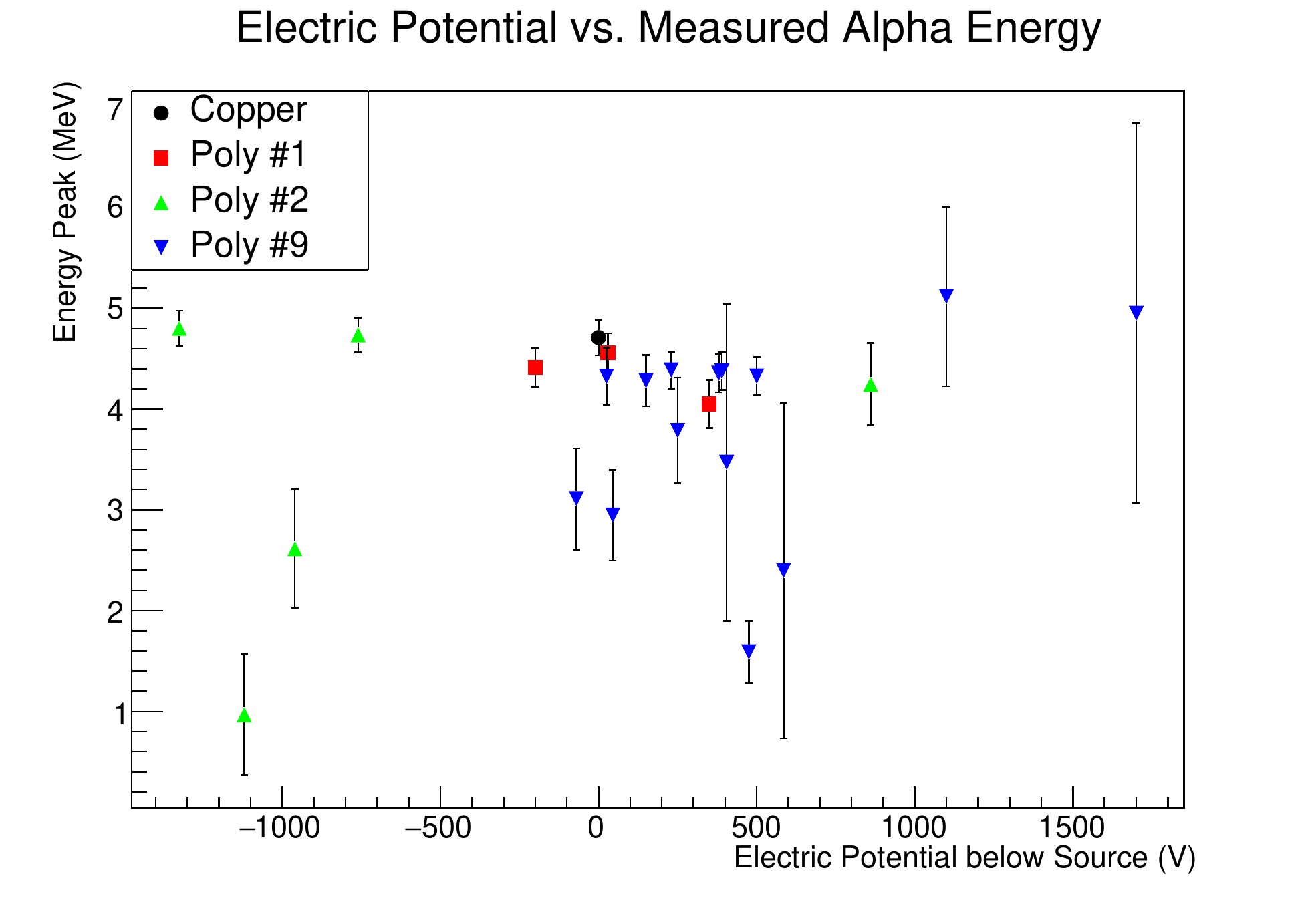}
    \includegraphics[width=0.9\columnwidth]{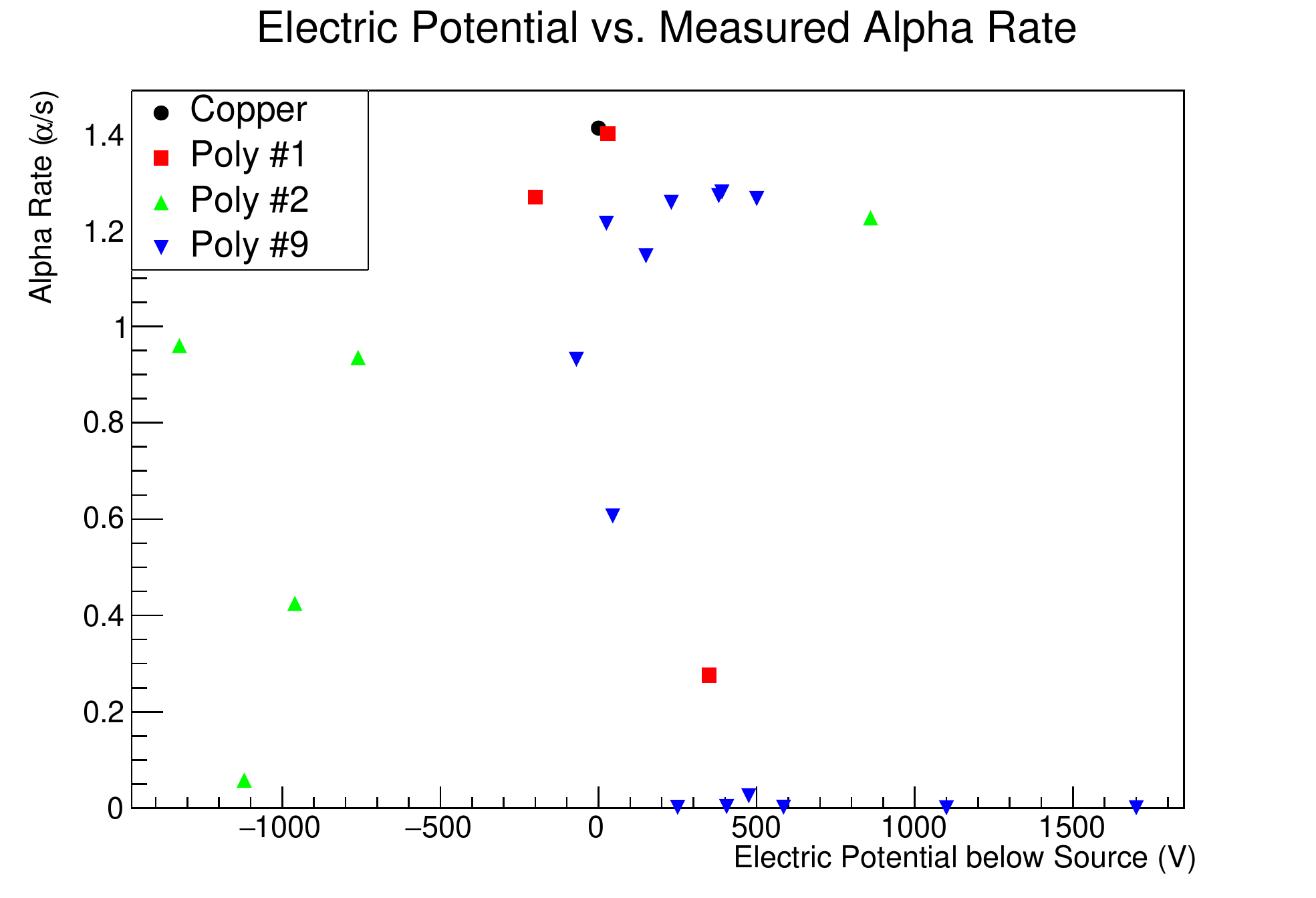}
    \caption{Electrostatic potential of polyethylene (poly) or copper directly below $^{230}$Th compared to XIA measurements of exiting alpha energy peak and rate. In cases where measurements were taken before and after the run, the average electric potential was taken.}
    \label{roughTrends}
\end{figure}

Compared against the copper control run, the pulse shape and energy characterization of runs with the $^{230}$Th placed on positive and negative regions of polyethylene were compared. As seen in Fig. \ref{roughTrends}, no clear correlation could be constructed based solely on the electrostatic potential beneath the source in regards to the measured energy peak and alpha production rate. 
Understanding the impact of embedded charge on detector response required a much higher fidelity understanding of the electric potential and associated field than a single point (or the rudimentary points of the ‘rough’ measurements).

\begin{figure}[ht]
    \centering
    \includegraphics[width=0.75\columnwidth]{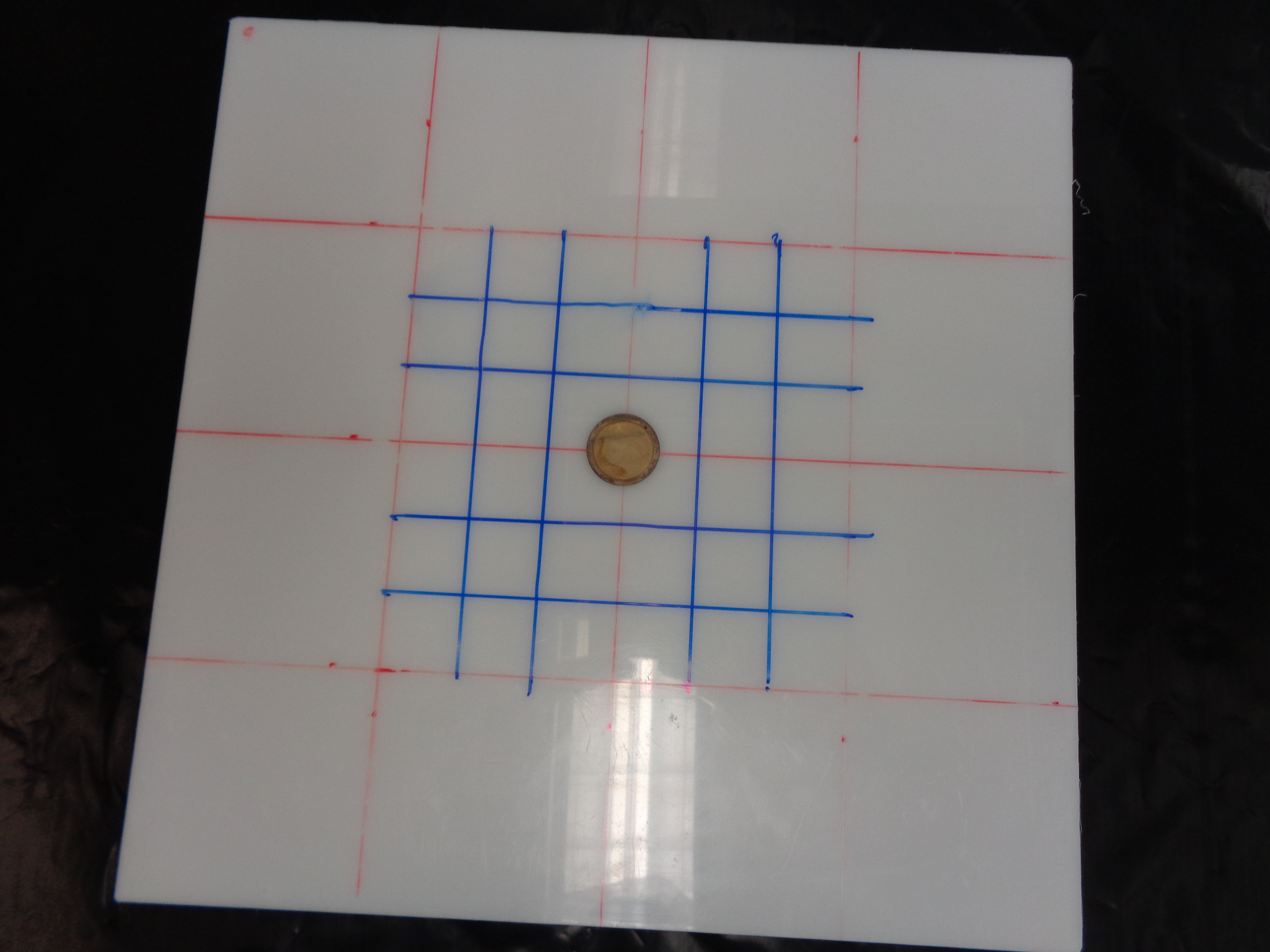}
    \caption{Photo of ‘fine’ XIA measurement setup with $^{230}$Th sample placed at center position. Blue lines mark 1 in intervals for polyethylene charge measurements.}
    \label{finesetup}
\end{figure}

\subsection{‘Fine’ Polyethylene Measurements}
\label{fine}
‘Fine’ runs kept the $^{230}$Th sample at the center position but took higher fidelity measurements of the potential from the polyethylene close to the center. 
In addition to measuring the nine points formed by the red markings, we took the forty points at 1 in intervals within the 6 in×6 in region centered on the source and the twelve points centered in the outermost squares formed by the red grid (see Fig. \ref{finesetup}). With the additional measurements of potential (and therefore longer duration of the tray kept open), substantially more outgassing was suspected to occur in subsequent runs, leading to an improper measurement of energy scale and alpha rate. For this reason, the polyethylene pieces were kept in the XIA for multiple 14-hour runs and the potential was measured as immediately as possible once completed.

The additional measurements allowed for the construction of a high fidelity map of the electric field within the chamber. 
FENiCSx was used to solve the Laplace equation for the potential with specified boundary conditions and take the gradient \cite{10.1145/1731022.1731030,10.1145/3524456,Scroggs2022}.
The boundary conditions, in this case, included the upper and lower anodes, field shapers, and measured potential on the polyethylene panels. For this calculation, the voltage of the upper anode was lowered from 1100 V to 1038.5 V, as this best fit the actual risetime (real electron drift being impacted by moisture and other environmental contamination).

%% \begin{figure}[ht]
%%     \centering
%%     \includegraphics[width=0.75\columnwidth]{figures/modeled_alpha.pdf}
%%     \caption{Fraction of all randomly distributed points for electric field calculations. Alpha modeled in blue and included nearby points in red.}
%%     \label{alphamodel}
%% \end{figure}

\begin{figure}[ht]
    \centering
    \includegraphics[width=0.9\columnwidth]{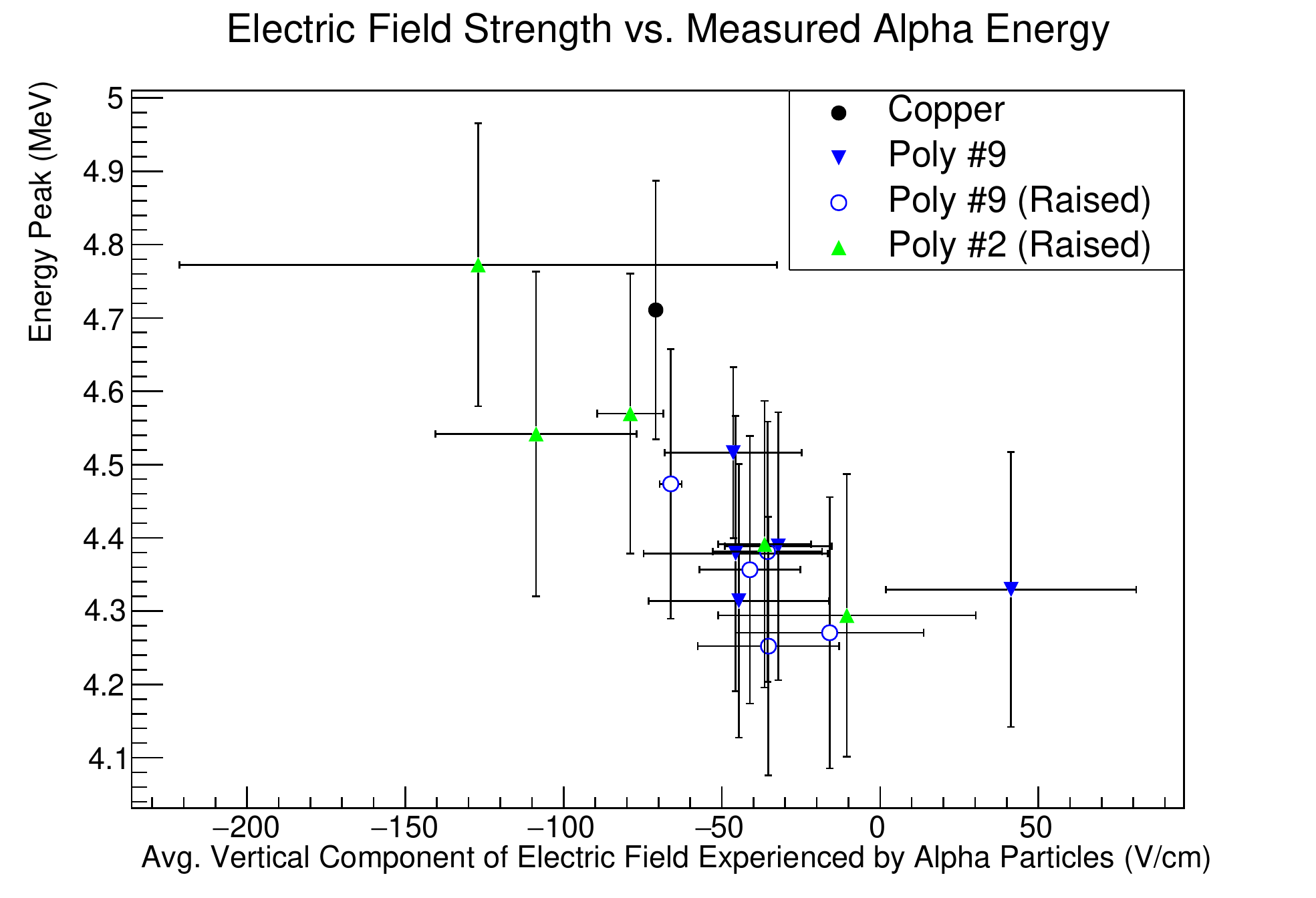}
    \includegraphics[width=0.9\columnwidth]{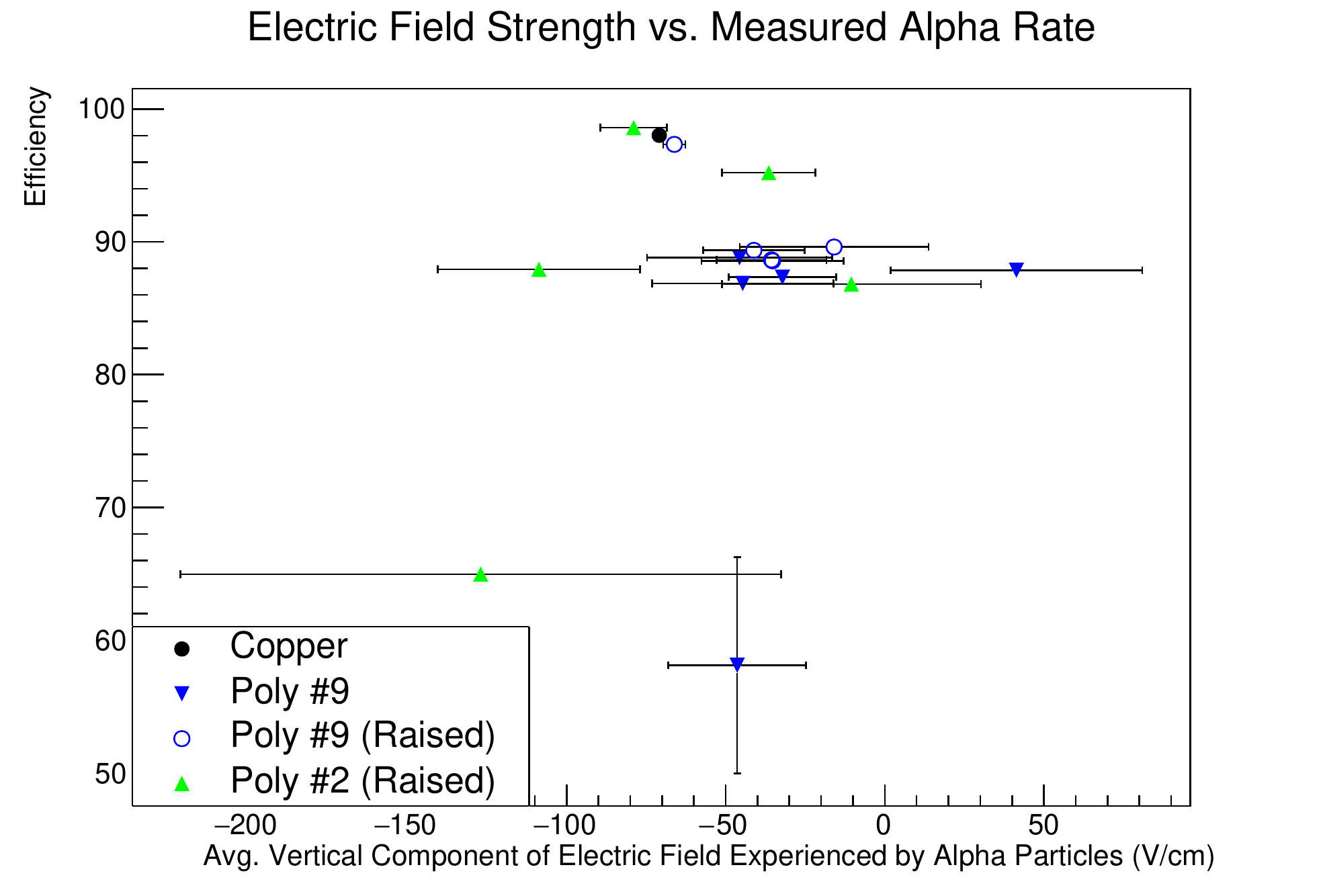}
    \caption{ Average vertical component of the electric field experienced by the average alpha particle compared to XIA measurements of exiting alpha energy peak and alpha rate efficiency. ‘Raised’ polyethylene was lifted from the grounding tray such that induced current was reduced, but the tray was lowered such that there was no actual change in height. Runs with significant outgassing (i.e. those with newly placed poly) not included.}
    \label{fineTrends}
\end{figure}

A sampling of $\sim$5,000 randomly distributed points was taken from a 4 cm hemisphere around the disc center within the chamber, each containing information pertaining to the electric field in its location. By modeling alphas as simple rays from the origin to the edge of the sphere, values from nearby points were summed and the average vertical component of the electric field for an alpha traveling in any direction from the source could be estimated. The mean of this value for all directions from the center was considered the average vertical component of the electric field experienced by the average alpha particle. It was used to correlate measurements in Fig. \ref{fineTrends}.

From three ‘fine’ polyethylene setups within the XIA, fifteen runs were measured and considered against the copper control. Within the $\sim$-70 V/cm to $\sim$0 V/cm  region in which the majority of polyethylene runs occurred, measurements for energy peak showed a consistent downwards trend in regard to the vertical electric field component (see Fig. \ref{fineTrends}). This follows from the understanding that positive electric potential leads to slowed (or even downwards) electron drift and absorption into the polyethylene sample rather than the upper anode (discussed further in Section \ref{Simulation}). The measured alpha efficiency showed no strong correlation. The majority of polyethylene runs maintained an $\sim$89\% alpha efficiency regardless of the electric field strength, with few outliers.

\begin{figure}[ht]
    \centering
    \includegraphics[width=0.75\columnwidth]{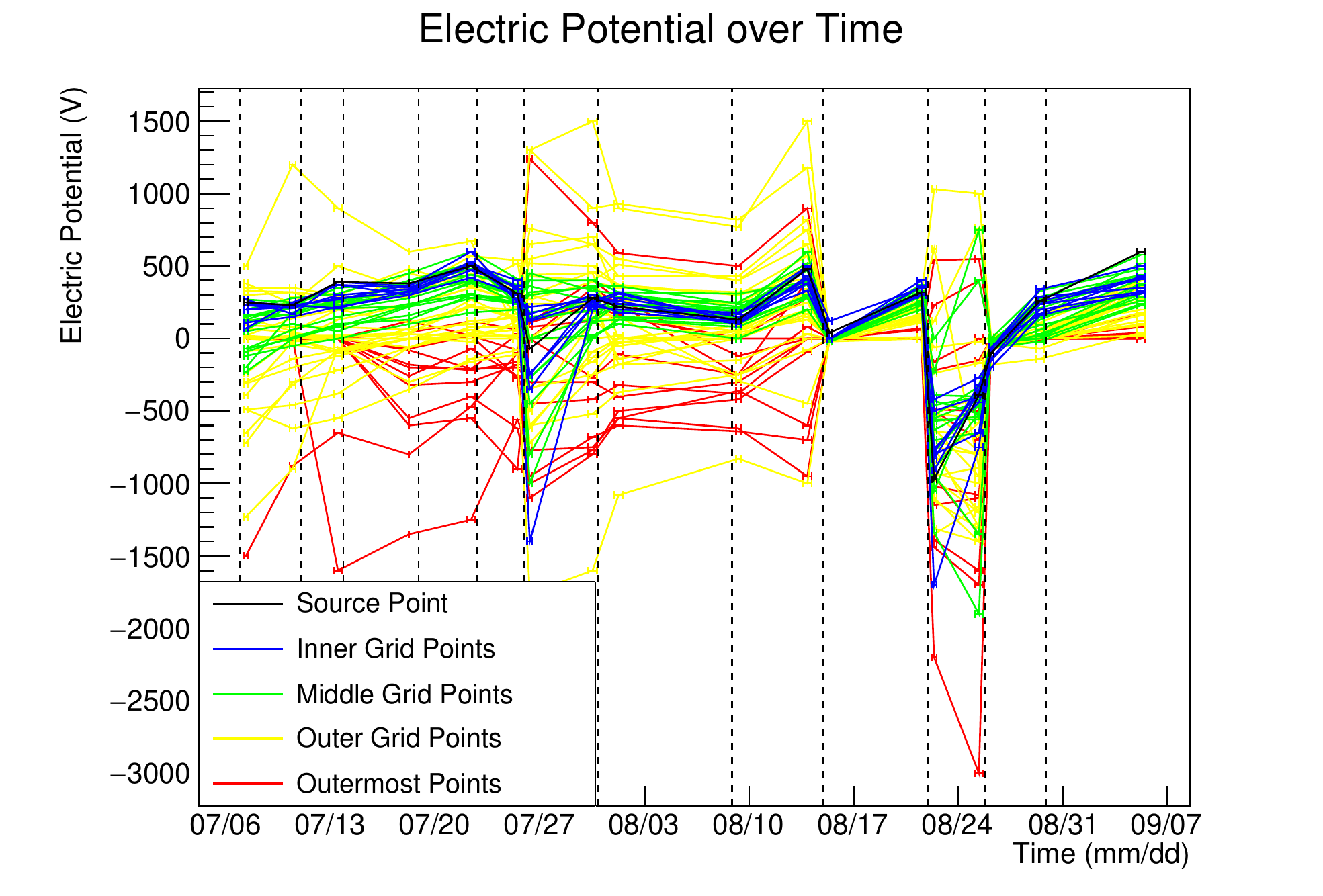}
    \caption{Measured electrostatic potential of points on polyethylene surface based on proximity to edge. Purges are marked with dotted vertical lines. }
    \label{chargedrift}
\end{figure}

Due to the relative consistency of the electric field strength measured across runs (and therefore small spread of data along the x-axis), a robust trendline or correlation in regard to energy peak cannot be constructed. Additionally, the significance of the outliers in alpha efficiency is unclear. Throughout the course of experimentation, we noticed frequent positive drift in the electric potential on the surface of the polyethylene plates, especially in the innermost region. Fig. \ref{chargedrift} shows the electrostatic potential of points across time. While the significant polyethylene manipulations (i.e. placing a new panel, raising a panel, or using an anti-static fan) at the purges on 7/7, 7/26, 8/15, 8/22, and 8/26 are somewhat obfuscating, electric potential measurements for the innermost points became increasingly positive and homogeneous in value over time, while measurements at outer points showed less consistent results. We performed an experiment with a polyethylene panel placed between a grounded table and a positive copper plate such that the electric field strength was similar to that within the XIA (a few hundred volts/day) and observed similar positive drift, suggesting that the drift was likely a result of the electric field, as discussed in Ref. \citenum{fleming_1999}. %rather than specific electronics in the XIA. 
More controlled experimentation is required to quantify this effect and achieve a wider range of measurements across the vertical electric field component for analysis.

\begin{figure}[ht]
    \centering
    \includegraphics[width=0.495\columnwidth]{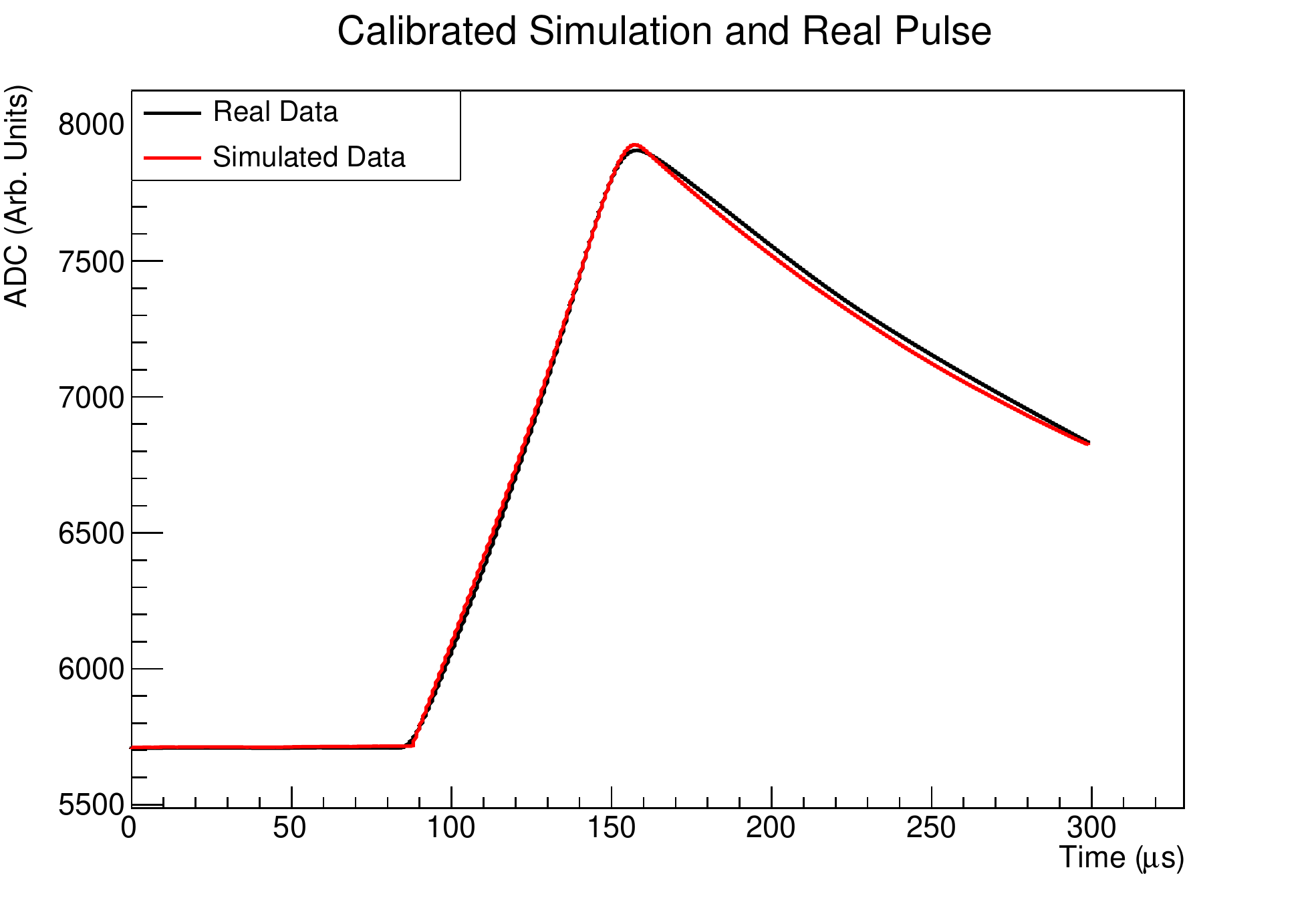}
    \includegraphics[width=0.495\columnwidth]{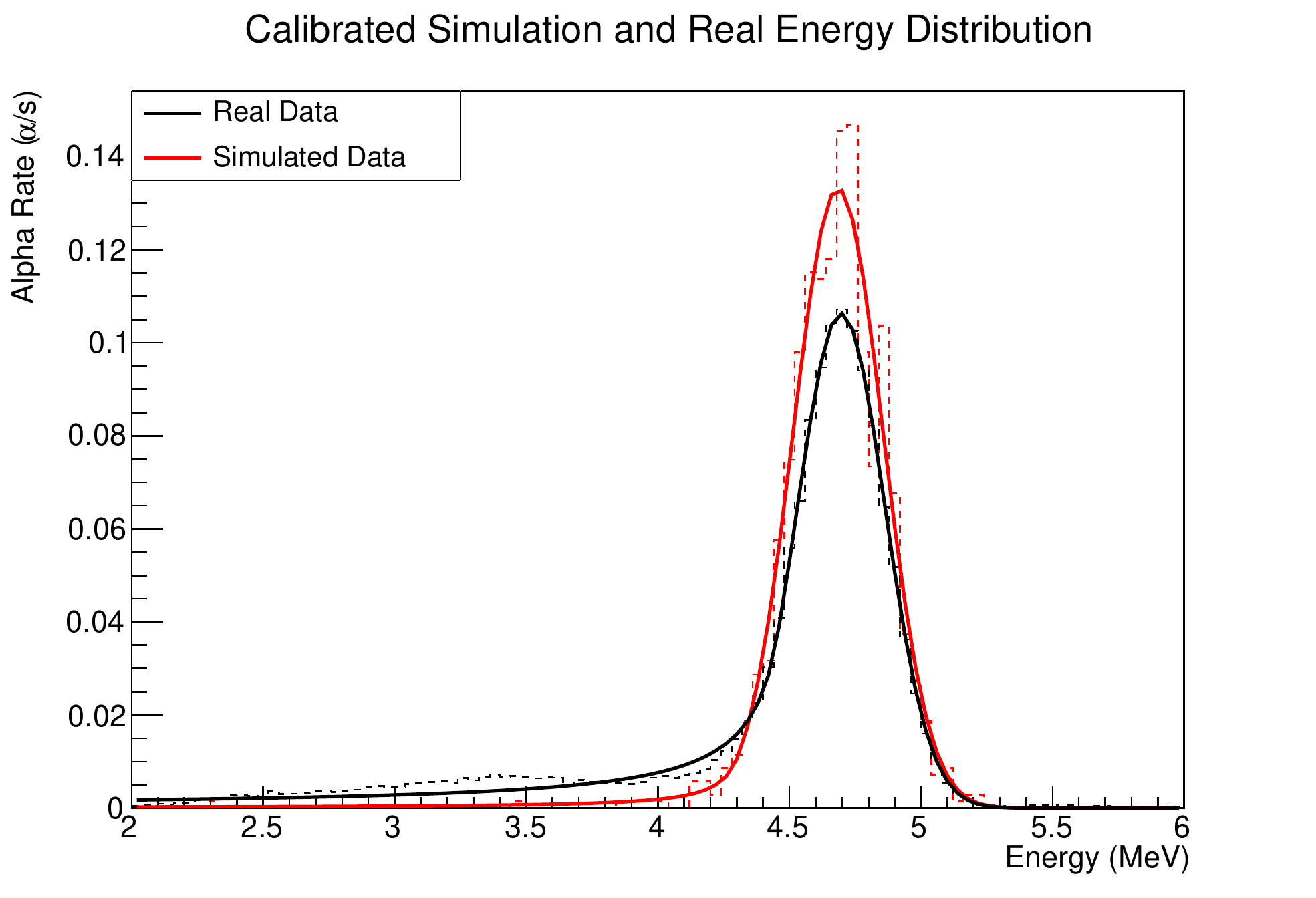}
    \caption{A comparison of the average pulse shape and energy distribution outputs of the XIA measurement with the $^{230}$Th source directly on the tray and the calibrated simulation.}
    \label{calibratedSim}
\end{figure}

\section{Simulation}
\label{Simulation}

To further investigate the discrepancies in pulse shape, exiting alpha energy, and alpha rate measured by the XIA as a result of extraneous charge, a simulation was used to predict electron motion. 
The Geant4 toolkit was used to model XIA geometry and the trajectory of alpha particles \cite{AGOSTINELLI2003250,ALLISON2016186,1610988}.\footnote{Code available at \url{https://bitbucket.org/rcalkins1/xia_ultralo1800_g4sim/src/master/}}
Garfield++ \cite{garfield}, HEED \cite{HEED} and Magboltz \cite{Magboltz} were used to implement electron drifts through the electric field (itself modeled as described in Section \ref{fine}). An integration of the induced current (as solved through the Shockley-Ramo theorem) alongside an exponential decay was used to simulate the electronics, as the precise components and values of the XIA's circuitry are proprietary and inaccessible \cite{posada}.

For the purposes of our analysis, the simulation was calibrated against measurements from the $^{230}$Th placed directly on the XIA tray (see Fig. \ref{calibratedSim}). It is important to note that the simulation did not account for the low energy tail present in real data and used a mono-energic alpha, making the energy distribution dominated by the peak amplitude and able be fit with a Gaussian function. The simulation was calibrated to reconstruct the real energy peak at 4.69 $\pm$ 0.17 MeV. 

\begin{figure}[ht]
    \centering
    \includegraphics[width=0.75\columnwidth]{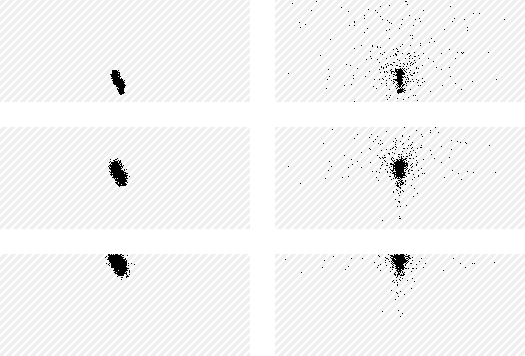}
    \caption{Diagram of simulated electrons from alpha after 15 $\mu$s (top), 40 $\mu$s (middle), and 65 $\mu$s (bottom) from runs with a copper control (left) and a largely positively charged polyethylene panel (right).  This view shows a projection of all electrons onto the sideplane of the XIA. }
    \label{electronSpread}
\end{figure}

\begin{figure}
    \centering
    \includegraphics[width=0.75\columnwidth]{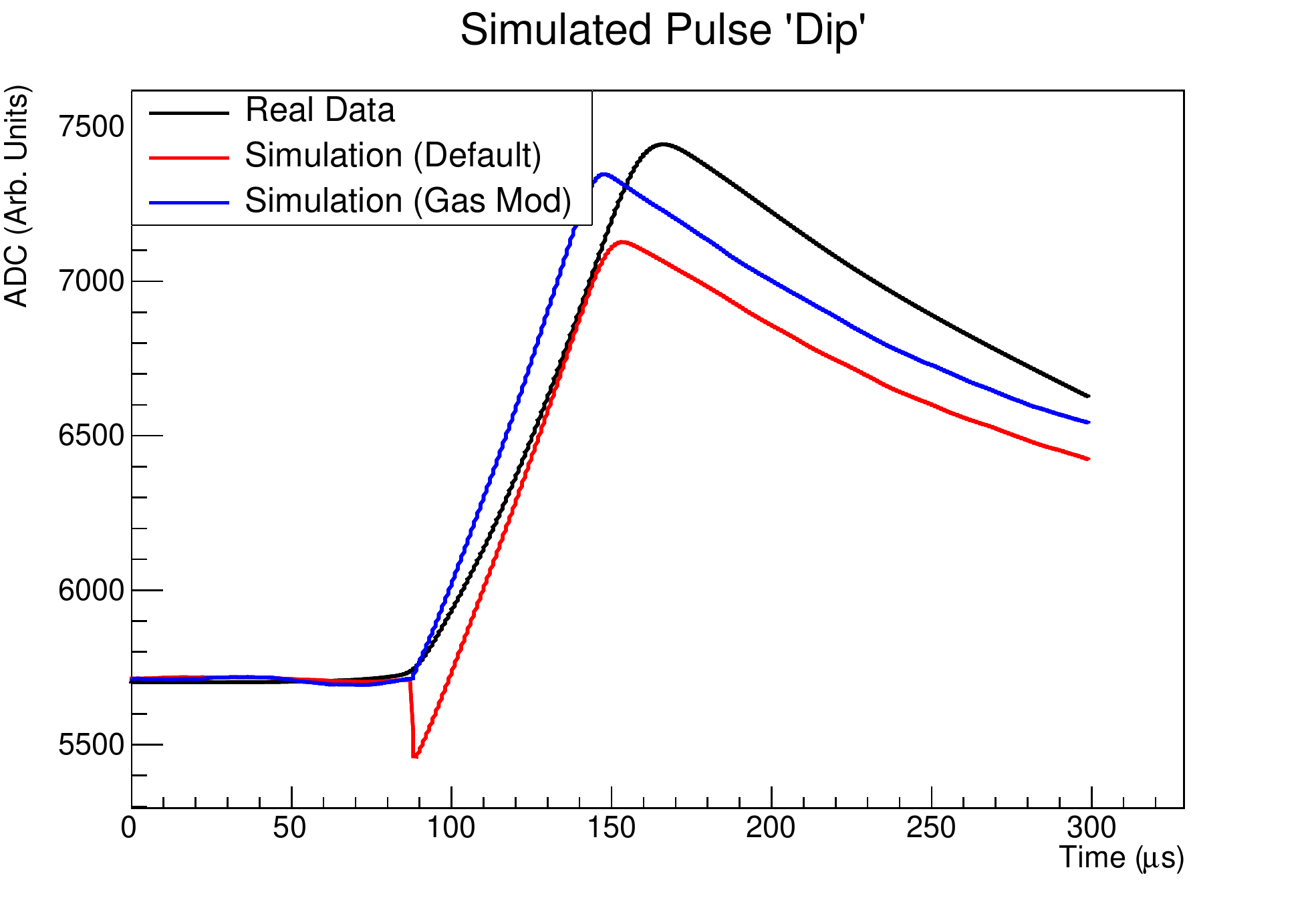}
    \caption{A comparison of the average pulse shape of an XIA measurement with positively charged polyethylene compared to the associated simulated pulse shapes.}
    \label{dip}
\end{figure}

\begin{figure}[ht]
    \centering
    \includegraphics[width=0.9\columnwidth]{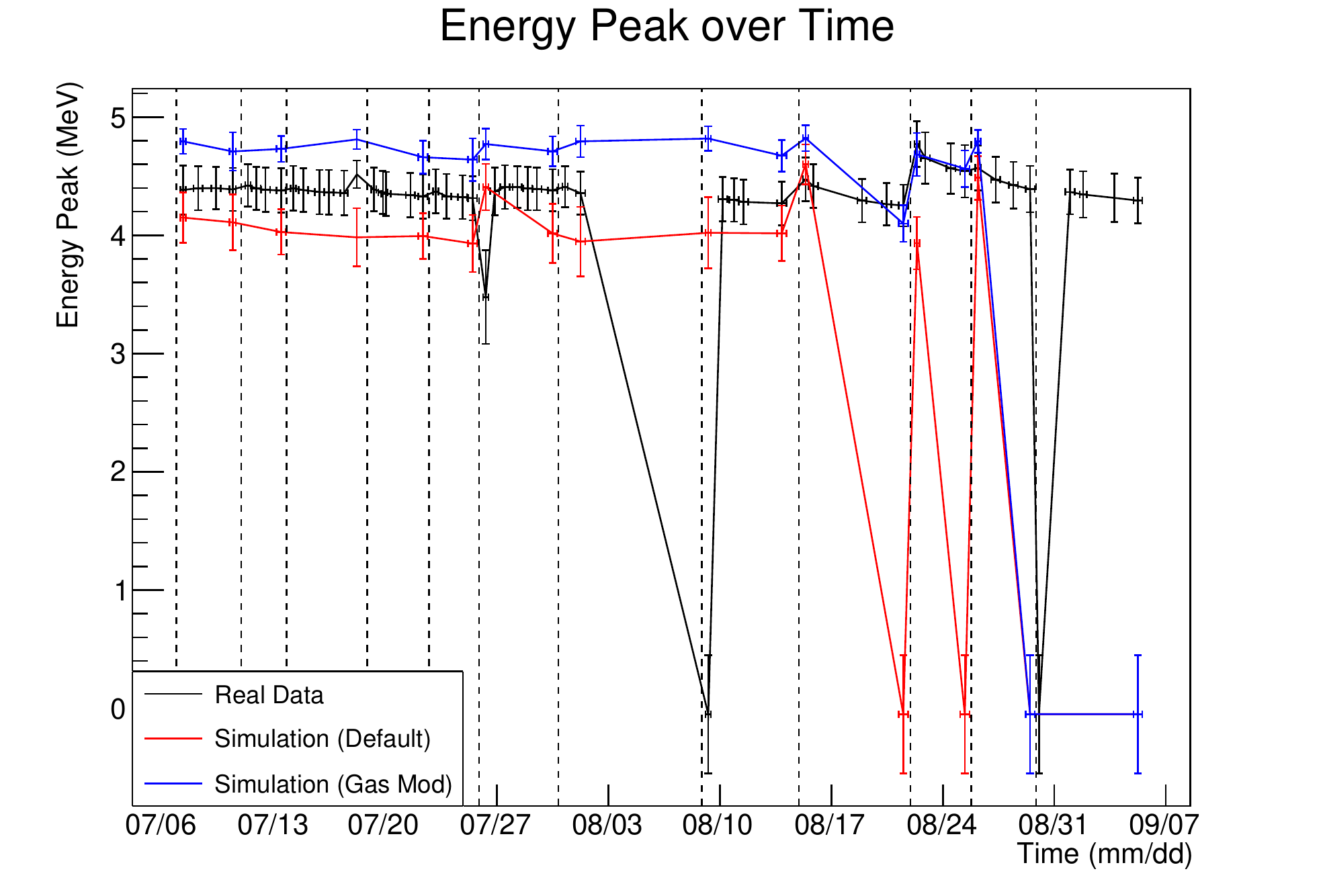}
    \includegraphics[width=0.9\columnwidth]{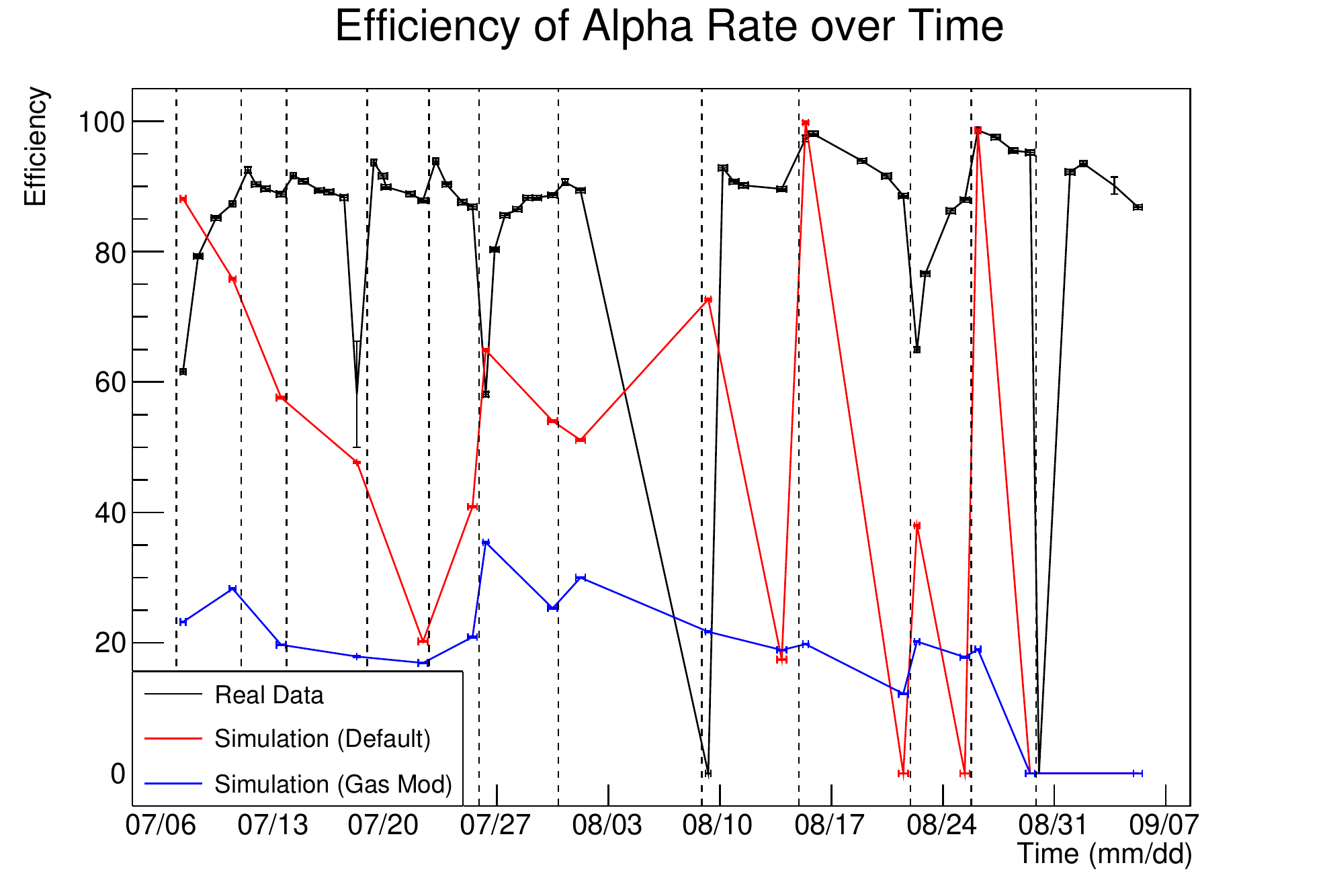}
    \caption{Comparisons of energy scale and alpha rate efficiency between real XIA data and simulations. Purges are marked with dotted vertical lines.}
    \label{simComparisons}
\end{figure}

As a consequence of the non-parallel electric field lines, we found that electron motion arising from runs with polyethylene panels had significant horizontal spread compared to runs with the copper control, as illustrated in Fig. \ref{electronSpread}. Electrons occasionally became ‘trapped’ above positively charged polyethylene, as regions in the chamber formed where they had zero vertical velocity due to the positive charge in the polyethylene and upper anode. Additionally, it was not uncommon for electrons to fail to reach the anode entirely as a result of the positive charge.

Simulated pulse shapes regularly involved negative induced charge, producing a characteristic ‘dip’ in ADC not present in real XIA data, as shown in Fig. \ref{dip}. As the characteristics of the pulse shape, such as the electron risetime (the duration of the initial positive slope in the pulse), are used to register events as either ‘alphas’ (particles originating from the radioactive sample) or ‘non-alphas’ (particles originating from counter sidewalls or a tray outside the sample), this phenomenon led to an improper categorization of events as ‘non-alpha’ particles and a low recorded alpha production rate. This is quantitatively shown as the ‘dip’ appearing in 49.0\% of all simulated events, but disproportionately in favor of particles not registered as alphas (appearing in 27.5\% of ‘alphas’ and in 79.2\% ‘non-alphas’). Typical XIA field strengths are approximately 70 V/cm, but charged samples with oppositely polarized points in close proximity can extend field strengths to hundreds of Volts per cm. By extending the Magboltz calculation of gas properties to higher electric field strengths and higher electrons velocities, the overall induction was positive (there was still some backscattering,  but not enough to significantly impact results). Simulated pulses under the extended gas file had modified properties which led to a quicker overall risetime, as shown in Fig. \ref{dip}.

Comparisons between ‘fine’ runs, the default simulation, and the simulation with the gas modification can be seen in Fig. \ref{simComparisons} in terms of the efficiency of alphas detected and the observed energy peak (as found from the Crystal Ball fit). The energy peak of the default simulation was 0.34 $\pm$ 0.23 MeV lower than real data and the simulation with the gas modification was 0.24 $\pm$ 0.18 MeV higher. In regards to alpha efficiency, the default simulation was 31 $\pm$ 24 \% lower than real data and the simulation with the gas modification was 65 $\pm$ 11 \% lower. These values do not include simulations where no alpha particles were registered (displayed in Fig. \ref{simComparisons} where both energy peak and alpha efficiency are zero). For such simulated runs, the electric potential of the polyethylene sample was positive across the surface without much variation (particularly around the source), potentially explaining why no pulses were registered. However, this is pure conjecture and does not explain why a similar effect was not observed in real data.

\section{Mitigation with Anti-Static Fan}
\label{Antistatic}
\subsection{Methodology}
Due to the significant influence of embedded charge on alpha measurements, we investigated the possibility of mitigation of the shown effects. Anti-static fans can be used to flood a polyethylene panel with positive and negative ions, thereby neutralizing the charge that had built up on its surface. As such, we utilized a commercially available anti-static fan in an attempt to eliminate the embedded charge and mitigate its influence on measurements.

\begin{figure}[ht]
    \centering
    \includegraphics[width=0.495\columnwidth]{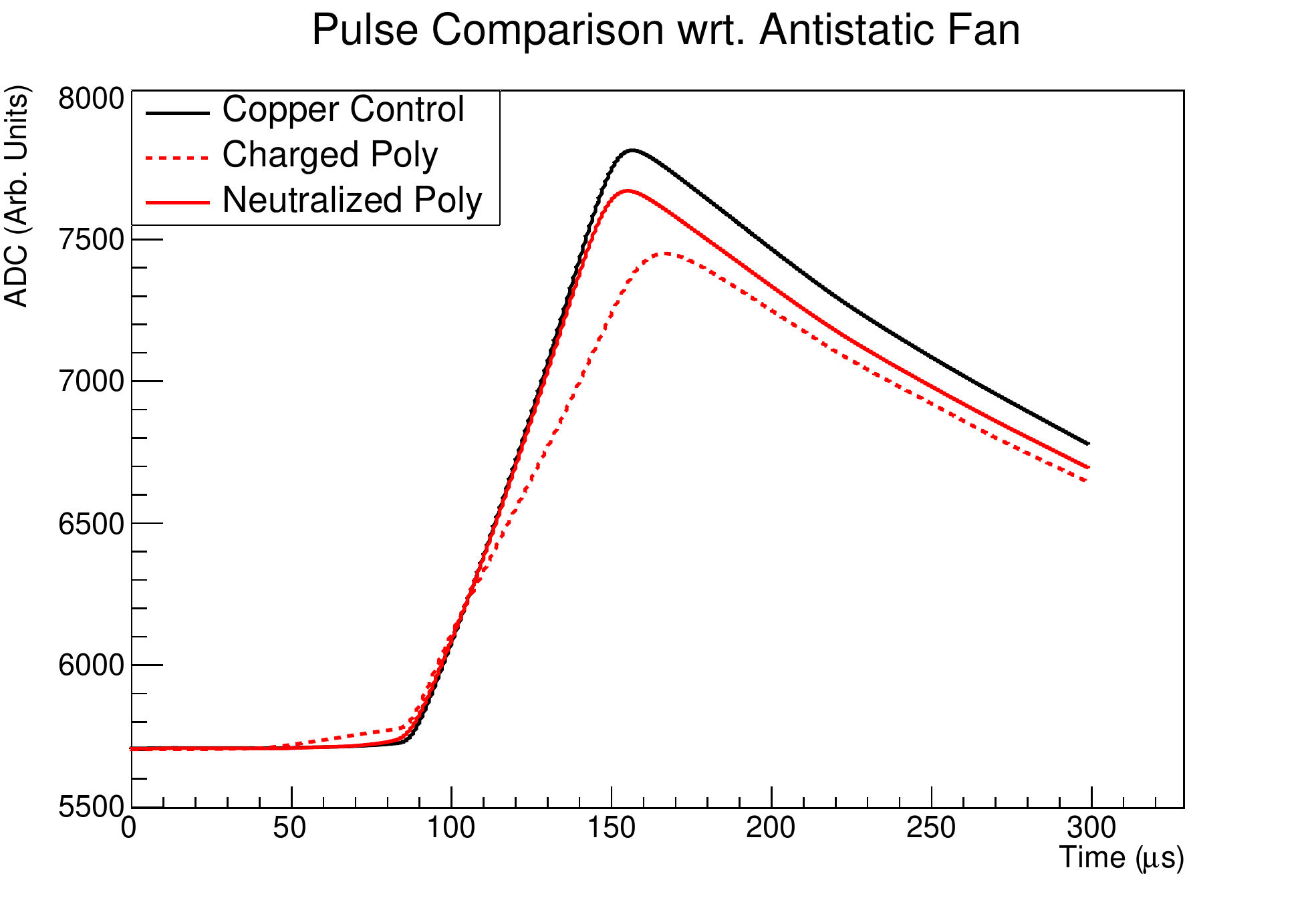}
    \includegraphics[width=0.495\columnwidth]{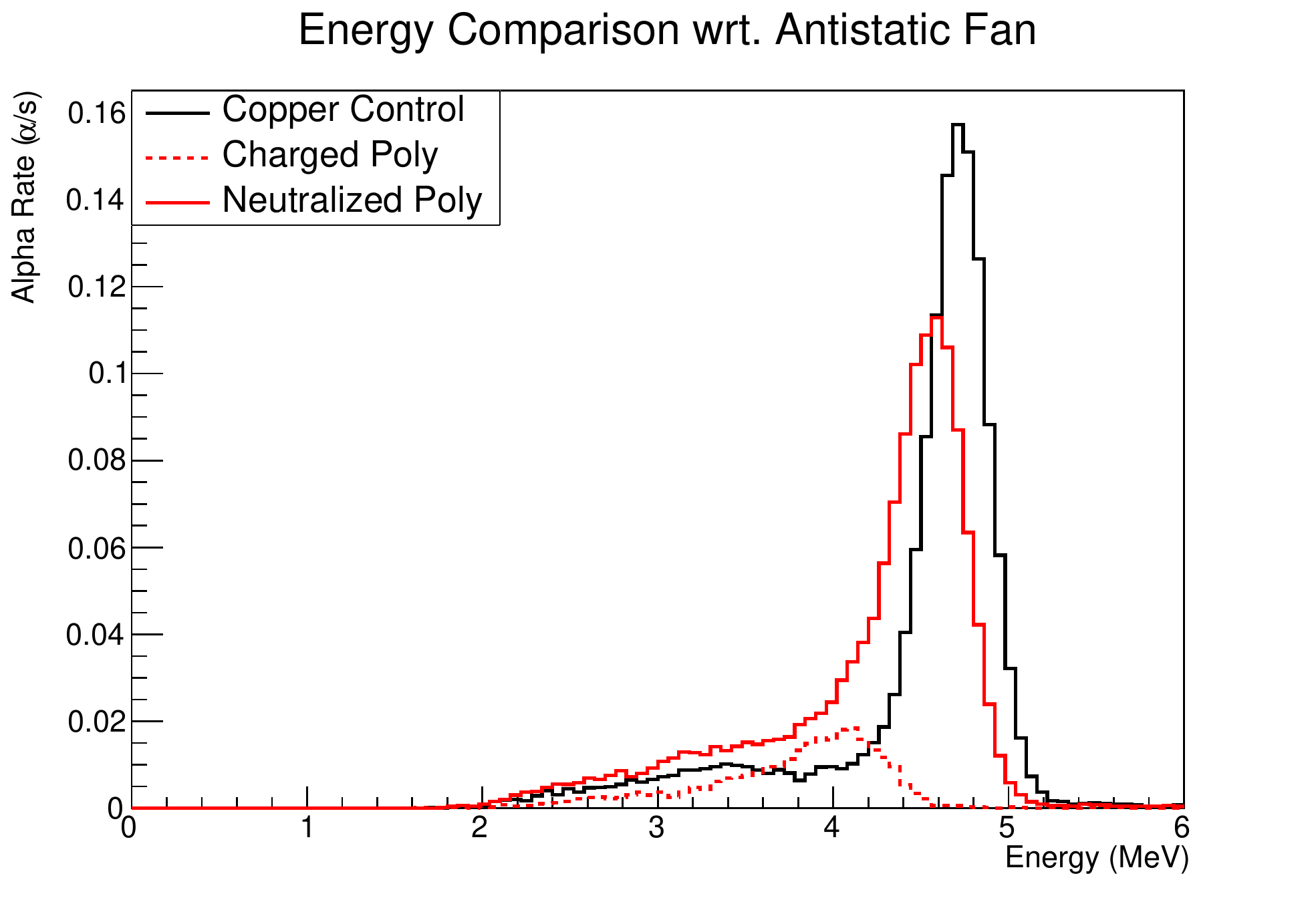}
    \caption{A comparison of the average pulse shapes and energy distributions from a polyethylene run (poly) before (charged) and after (neutralized) the use of an anti-static fan compared to those from the copper control.}
    \label{antistaticcompare}
\end{figure}

Similar to prior procedures, polyethylene panels of dimensions 24 in×24 in×3/16 in were prepared on the XIA tray with a centered $^{230}$Th source. Electrostatic potential measurements were taken at the points marked by the grids on the polyethylene. Following a 14-hour measurement period in the XIA, the anti-static fan was used over the full polyethylene front and back surfaces, measurements of the electrostatic potential were taken, and the sample was re-entered into a 14-hour run.

\subsection{Results}
While the values for electric potential measured on the polyethylene prior to the anti-static fan were not the most extreme of those observed (nor were they completely neutralized after the anti-static fan), there were reductions in the magnitude of electric potential of 65.93\%, 97.89\%, and 94.29\% across the panels (the latter two can be seen at 8/15 and 8/26 in Fig. \ref{chargedrift}). In comparing the average pulse shapes of one polyethylene experiment against the copper control, as seen in Fig. \ref{antistaticcompare}, it is qualitatively clear that measurements following the anti-static fan (‘neutralized’) were much nearer to the calibrated copper control measurements compared to those prior to use of the anti-static fan (‘charged’). The energy distributions similarly corroborate that the anti-static fan brought measurements much closer to the expected control.

\begin{table}[ht]
    \centering
    \begin{tabular}{c | c | c | c | c}
        \multicolumn{5}{c}{Alpha Measurements from Anti-Static Fan} \\

        \hline \hline
                       & Energy & Alpha      & Electron & Pulse     \\
                       & Peak   & Efficiency & Risetime & Amplitude \\
                       & (MeV)  & (\%)       & ($\mu$s) & (ADC)     \\
        \hline \hline 
        Copper Control & 4.71 $\pm$ 0.18 & 98.02 $\pm$ 0.49 & 70.0 $\pm$ 3.6 & 7813 $\pm$ 288 \\
        \hline
        \underline{Poly \#1} & & & & \\
        Charged        & 4.05 $\pm$ 0.24 & 19.18 $\pm$ 0.38 & 81.4 $\pm$ 2.7 & 7450 $\pm$ 222 \\
        Neutralized    & 4.56 $\pm$ 0.19 & 97.24 $\pm$ 0.48 & 69.7 $\pm$ 3.8 & 7671 $\pm$ 283 \\
        Subsequent     & 4.41 $\pm$ 0.19 & 88.08 $\pm$ 0.35 & 75.5 $\pm$ 3.9 & 7570 $\pm$ 287 \\
        \hline
        \underline{Poly \#9 (Raised)} & & & & \\
        Charged        & 4.27 $\pm$ 0.18 & 89.61 $\pm$ 0.35 & 77.3 $\pm$ 3.9 & 7450 $\pm$ 397 \\
        Neutralized    & 4.47 $\pm$ 0.18 & 97.34 $\pm$ 0.48 & 67.9 $\pm$ 3.6 & 7580 $\pm$ 355 \\
        Subsequent     & 4.25 $\pm$ 0.17 & 88.57 $\pm$ 0.35 & 78.3 $\pm$ 3.6 & 7481 $\pm$ 357 \\
        \hline
        \underline{Poly \#2 (Raised)} & & & & \\
        Charged        & 4.54 $\pm$ 0.22 & 87.93 $\pm$ 0.35 & 67.8 $\pm$ 4.1 & 7483 $\pm$ 427 \\
        Neutralized    & 4.57 $\pm$ 0.19 & 98.60 $\pm$ 0.49 & 67.9 $\pm$ 3.7 & 7644 $\pm$ 345 \\
        Subsequent     & 4.39 $\pm$ 0.20 & 95.21 $\pm$ 0.36 & 74.4 $\pm$ 4.0 & 7533 $\pm$ 371 \\
        \hline \hline
	
    \end{tabular}
    \caption{Table of alpha particle measurements from polyethylene runs (poly) before (charged), immediately after (neutralized), and some time later (subsequent) using an anti-static fan compared to those from the copper control.}
    \label{antistatictable}
\end{table}

Tbl. \ref{antistatictable} provides a quantitative look at the primary parameters analyzed for alpha particle measurements. To varying degrees of effectivity, the use of an anti-static fan proved a practical solution to mitigating the effects of charged polyethylene (rather than attempting to simulate its complex influence). The percent error of measurements of the characteristics of the average pulse shape, electron risetime and pulse amplitude, were brought to within 2.09\% and 2.33\%, respectively, of those from the copper control. Measurements of energy peak and alpha efficiency were brought to within 3.74\%  and 0.695\%, respectively. However, as a result of the charge drift discussed in Sec. \ref{fine}, measurements following neutralization drifted away from the copper control after some time.

\section{Conclusion and Outlook}
After recommissioning the XIA UltraLo-1800 Alpha Particle Counter, we have shown that the XIA can be successfully stored with N$_2$ boil-off gas and be expected to maintain calibration even following a prolonged time of inactivity.

Based on alpha particle measurements from the XIA with a $^{230}$Th source and polyethylene panels with embedded charges, we conclude that the static charges present in non-conductors is likely a primary factor influencing the variable rates of $^{210}$Pb accumulation on polyethylene in SNOLAB or at the very least, an influencing factor on the assays. There was no obvious correlation between the measurements of alpha particles and the electric potential beneath the radioactive sample, but one did exist for measured alpha energy peak and the average vertical component of the electric field experienced by the average alpha particle, at least within the $\sim$-70 V/cm to $\sim$0 V/cm region. The existence of a correlation for alpha detection efficiency remains unclear.

We found a predictive model of the alpha particles and free electrons resulting from the ionization track in the gas can be used, assuming the form of the nearby electric field is sufficiently understood, but further studies are necessary to fully model the extent of environmental influences and effects of XIA circuitry. There are significant issues in modeling that must be addressed prior to use as a tool for correcting alpha measurements.
Additionally, the use of an anti-static fan to neutralize the charges on the polyethylene serves as an appropriate method of mitigating the effects of the static charges, thereby recovering 97.73\% alpha detection efficiency. In practice, as the effects of the anti-static fan disappear after a few days, this method would require regular use on the sample to keep a consistently neutralized charge. The development of charge on polyethylene is further discussed in Refs. \citenum{fleming_1999} and \citenum{GROMOV1989405}.

Further research could involve finding the maximum electric potential gradient possible on the polyethylene and measuring the panels at significantly small intervals to best chart the electric potential for a more robust correlation and simulation. Such a study may be able to correct for the influence of the static charges through a revised simulation, allowing it to be used as a tool for analyzers to ‘correct’ future work and alpha particle measurements.

\section*{Acknowledgements}
We would like to thank Professor Jodi Cooley for her support and guidance throughout the course of this research. We would like to thank the SuperCDMS Backgrounds Working Group for their advice and feedback. This material is based upon work supported by the National Science Foundation under Grant No. 2111457. Any opinions, findings, and conclusions or recommendations expressed in this material are those of the author(s) and do not necessarily reflect the views of the National Science Foundation. 

Additionally, we would like to thank the Southern Methodist University (SMU) Office of Engaged Learning and Summer Research Intensive program for their funding and support over the course of this experiment. We would like to thank SMU and the SMU Physics Department and administration for their continued support and foundation of education. We are grateful to the leadership and staff of SMU's high performance computer cluster ManeFrame II, which allowed us to perform the simulations necessary for analysis.



%\bibliographystyle{elsarticle-num} 
\bibliographystyle{JHEP}
\bibliography{references}

\end{document}